\documentclass[sigplan,nonacm,pbalance]{acmart}
\AtBeginDocument{%
  }

\setcopyright{acmlicensed}
\copyrightyear{2018}
\acmYear{2018}
\acmDOI{XXXXXXX.XXXXXXX}
\acmConference[Conference acronym 'XX]{Make sure to enter the correct
  conference title from your rights confirmation email}{June 03--05,
  2018}{Woodstock, NY}
\acmISBN{978-1-4503-XXXX-X/2018/06}

\usepackage[ruled, lined,linesnumbered, longend]{algorithm2e}
\usepackage{tikz}
\usetikzlibrary{tikzmark}
\usepackage{stfloats}
\usepackage{threeparttable}
\usepackage{multirow}
\usepackage{graphicx}
\usepackage{xcolor}
\usepackage{listings}
\usepackage{booktabs}
\usepackage{subcaption}
\usepackage[frozencache,cachedir=mcache]{minted}
\usepackage{siunitx}

\newif\ifdraft %
 \drafttrue

\newif\ifreview
\makeatletter
  \if@ACM@anonymous
    \reviewtrue
  \else
    \reviewfalse
  \fi
\makeatother

\newcommand{\defname}[3]{
    \ifreview
        \newcommand{#1}{#3}
    \else
        \newcommand{#1}{#2}
    \fi
}

\defname{\MSAGA}{FSA}{ASA}
\defname{\msaga}{fsa}{asa}
\defname{\SA}{SystolicAttention}{FusedAttention}
\defname{\MSAGAURL}{
    \url{https://github.com/VCA-EPFL/FSA}
}{
    \url{https://anonymous.4open.science/r/ASA}
}

\newcommand{\FA}{FlashAttention}
\newcommand{\FAthree}{FlashAttention-3}
\newcommand{\FLOPs}{FLOPs/s}
\newcommand{\TFLOPs}{TFLOPs/s}
\newcommand{\FuseMax}{FuseMax}

\newcommand*\circled[1]{\tikz[baseline=(char.base)]{\node[shape=circle,draw,inner sep=1pt](char){#1};}}

\definecolor{SoftBlue}{RGB}{230,240,255}
\definecolor{SoftRed}{RGB}{255,235,235}
\definecolor{peachpuff}{RGB}{255,218,185}
\definecolor{softpeach}{RGB}{255,236,220}
\definecolor{SkyBlue}{rgb}{0.53, 0.81, 0.92}
\definecolor{LightSalmon}{rgb}{1.0, 0.63, 0.48}
\definecolor{RoyalBlue}{rgb}{0.25, 0.41, 0.88}
\definecolor{OrangeRed}{rgb}{1.0, 0.27, 0.0}

\newcommand{\tikzHighLight}[4]{
    \tikz[overlay, remember picture]{
        \fill[#2, opacity=0.3]
        ([xshift={#3[0]}, yshift={#3[1]}]pic cs:#1-start) rectangle ([xshift={#4[0]}, yshift={#4[1]}]pic cs:#1-end);
    }
}

\definecolor{typecolor}{rgb}{0.2,0.2,0.7}  %

\makeatletter
\@ifpackagelater{minted}{2024/10/01}{
    \newminted[mintedpython]{python}{
        extrakeywordstype={MTile,ATile,STile},
        fontsize=\small,
        linenos
    }
}{
    \newminted[mintedpython]{python}{
        fontsize=\small,
        linenos
    }
}
\makeatother
\newminted[mintedscala]{scala}{
    fontsize=\small,
    linenos
}

\begin{document}

\ifreview
\title[]{\MSAGA{}: Fusing \FA{} \\ within a Single Systolic Array}
\else
\title[]{\SA{}: Fusing \FA{} within a Single Systolic Array}
\fi

\author{Jiawei Lin}
\email{jiawei.lin@epfl.ch}
\affiliation{%
  \institution{EPFL}
  \city{Lausanne}
  \country{Switzerland}
}

\author{Yuanlong Li}
\email{yuanlong.li@epfl.ch}
\affiliation{%
  \institution{EPFL}
  \city{Lausanne}
  \country{Switzerland}
}

\author{Guokai Chen}
\email{guokai.chen@epfl.ch}
\affiliation{%
  \institution{EPFL}
  \city{Lausanne}
  \country{Switzerland}
}

\author{Thomas Bourgeat}
\email{thomas.bourgeat@epfl.ch}
\affiliation{%
  \institution{EPFL}
  \city{Lausanne}
  \country{Switzerland}
}

\renewcommand{\shortauthors}{Jiawei et al.}

\begin{abstract}

Transformer models rely heavily on the scaled dot-product attention (SDPA) operation,
typically implemented as \FA{}.
Characterized by its frequent interleaving of matrix multiplications and softmax operations,
\FA{} fails to fully utilize the compute resources of modern systolic-array-based accelerators
designed for consecutive and large matrix multiplications.

To fully unleash the performance potential of systolic arrays for \FA{},
we propose \MSAGA{}, an enhanced systolic array architecture that runs the entire \FA{}
on the array without external vector units.
Combined with \SA{}, an optimized kernel for \MSAGA{} that achieves fine-grained and element-wise
overlapping of \FA{} operations, \MSAGA{} maximizes array utilization while preserving the
original floating-point operation order of \FA{}.
We implement \MSAGA{} in synthesizable RTL and evaluate its performance against state-of-the-art
systolic-array-based accelerators.
Our results show that \MSAGA{} achieves 1.77$\times$ and 4.83$\times$ higher attention \FLOPs{} utilization
compared to AWS Neuron-v2 and Google TPUv5e, respectively.
We synthesize \MSAGA{} in a 16\,nm technology at 1.5\,GHz, and results indicate
only a 12\% area overhead compared to a standard weight-stationary systolic array.
\end{abstract}

\maketitle

\section{Introduction}

The recent rapid advancement of artificial intelligence is largely driven by the success of
Transformer models~\cite{attention-is-all-you-need,bert,gpt3}, which revolutionize applications
ranging from language translation, text generation~\cite{chatgpt}, image recognition~\cite{vit},
to autonomous driving~\cite{transformer-autonomous}.
Due to their enormous computational demands~\cite{energy-scaling-laws},
Transformer models are typically run on specialized hardware accelerators to
achieve practical performance and energy efficiency.
As these models continue to scale in size and complexity~\cite{scaling-laws},
the need for efficient hardware acceleration becomes increasingly critical.

\begin{figure}[t]
\includegraphics[width=1.0\columnwidth]{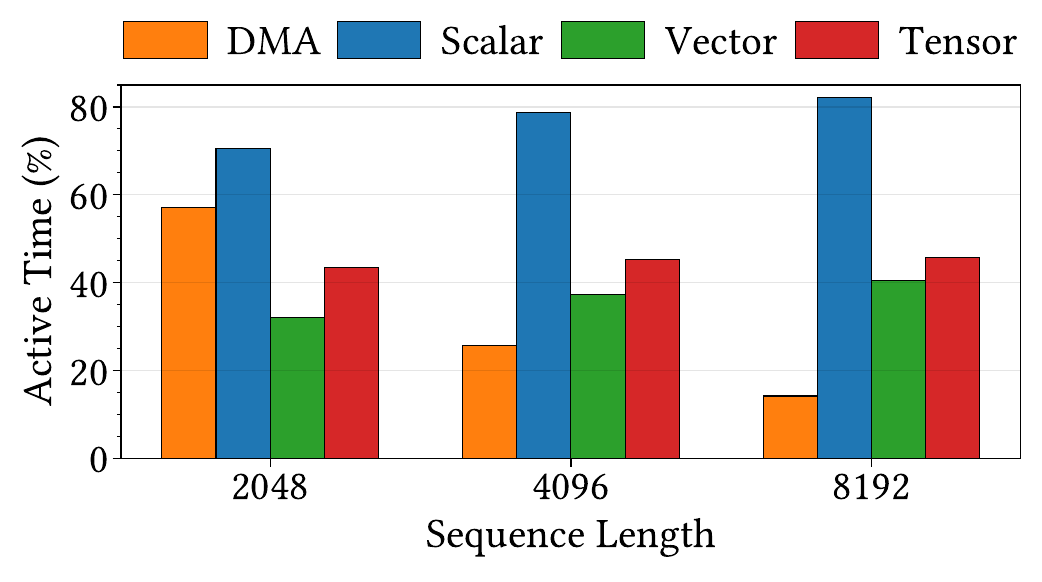}
\caption{Percentage of active time of various components in AWS NeuronCore-v2 when running \FA{}.}
\label{fig:aws-active-time}
\Description{Active Time}
\end{figure}

Systolic arrays~\cite{original_systolic} are widely used in both industry~\cite{google-tpu-arch,tpuv1,tpuv4,trainium-arch} and
academic accelerators~\cite{ucb-gemmini} to run deep learning workloads.
In systolic arrays, data is continuously streamed through a network of tightly coupled processing elements (PEs) to
maximize computation density while minimizing data movement overhead.
Despite various dataflows~\cite{eyeriss-jssc,eyeriss-isca,flat,fusemax} proposed for systolic arrays,
the multiply-accumulate (MAC)-based two-dimensional (2D) systolic array is most widely adopted,
as in Google's TPU~\cite{google-tpu-arch,tpuv1,tpuv4} and AWS NeuronCore~\cite{trainium-arch} that are mostly deployed for running Transformer models.

The key operation in Transformer models is the \emph{scaled dot-product attention} (SDPA),
which is usually implemented using the \emph{\FA{}}~\cite{fa2,fa3} algorithm.
When running \FA{} on systolic-array-based accelerators, matrix multiplications are executed on the systolic array,
while softmax operations are offloaded to external vector and scalar units.
The reliance on this simplistic division of labor forces vector/scalar units to supply adequate
floating-point operations per second (\FLOPs{}) for softmax,
failing to do so causes them to bottleneck the overall pipeline.
As shown in \autoref{fig:aws-active-time}, when running \FA{} on AWS NeuronCore-v2,
the systolic array (tensor engine) is only active for about 45\% of the time, while the scalar units remain 
active for 80\% of the time.

The simplistic mapping of computation also creates another source of low hardware utilization:
as \FA{} computes softmax between two matrix multiplications, data must be transferred
back and forth frequently between the systolic array and vector/scalar units through a local SRAM,
thereby preventing reuse of any matrix data across iterations,
increasing preloading overhead, and exacerbating SRAM port contention~\cite{virgo,mosaic}.
As shown in our experiments in \autoref{sec:eval-performance}, the 
systolic array of AWS NeuronCore-v2 only achieves 25\% of the
theoretical \FLOPs{} utilization, even though it is active for 45\% of the time.

We argue that the above performance problems can be addressed by running the entire \FA{}
solely on the systolic array, where matrix multiplications and softmax can utilize and share the systolic array \FLOPs{} at a finer granularity.
As a result, softmax is no longer limited by the throughput of vector/scalar units, and both
data movement overhead and SRAM port contention are eliminated as all computations occur at the same place.
To efficiently implement this approach, apart from executing softmax operations, including \texttt{exp}, \texttt{rowmax}, and \texttt{rowsum},
using only the systolic array, various operations must also be effectively overlapped to maximize array utilization.
FuseMax~\cite{fusemax} pioneered the efficient fusion of \FA{} on systolic arrays.
While it addresses the challenge of handling softmax by mapping \texttt{rowmax} and \texttt{rowsum} as spatial reductions,
its core strategy for achieving high hardware utilization is to process multiple \FA{} iterations simultaneously.
This parallel processing approach, however, necessitates storing multiple contexts inside each PE to enable context switching
among iterations, leading to additional hardware overhead.

We propose \emph{\MSAGA{}}, a new hardware-friendly systolic array architecture that
fully unleashes the performance potential of systolic arrays by running the entire \FA{}
on a single systolic array with minimal area overhead.
\MSAGA{} regards \texttt{rowmax} and \texttt{rowsum} as reduction operations and
performs them on-the-fly and in-place with an array of comparators and a specialized upward data path, thereby eliminating the
need to store them in local SRAM.
To calculate \texttt{exp}, \MSAGA{} reuses the systolic array to perform linear interpolation~\cite{exp_pwl}
to approximate the result, taking advantage of the fact that the input of \texttt{exp} is always less than or equal to zero.
Based on \MSAGA{}, we also introduce \emph{\SA{}},
an optimized kernel that makes the best use of \MSAGA{}'s capabilities to overlap
multiple operations within the same \FA{} iteration,
further improving array utilization while preserving the
order of all floating-point operations in \FA{}.

We make the following contributions in this paper:
\begin{itemize}
    \item
    We propose \MSAGA{}, an enhanced systolic array architecture that can execute the
    entire \FA{} solely on the systolic array without any vector or scalar units.
    We implement \MSAGA{} in synthesizable RTL, achieving clock frequency of 1.5\,GHz
    under 16\,nm technology.
    Our synthesis result shows that \MSAGA{} only incurs 12\% additional area overhead
    compared to the baseline systolic array.

    \item 
    We propose \SA{}, an optimized kernel for \MSAGA{} to efficiently overlap operations inside a single \FA{} iteration.
    With \SA{}, \MSAGA{} achieves 1.77$\times$ and 4.83$\times$ higher \FLOPs{} utilization
    than AWS NeuronCore-v2 and TPUv5e, respectively.

    \item
    We develop a Python programming interface that allows users to write custom kernels for \MSAGA{}.
    We open-source both the RTL implementation and the Python software stack at \MSAGAURL{}.
\end{itemize}

\section{Background and Motivation}

In this section, we first briefly introduce baseline systolic arrays and how
they perform matrix multiplications. We then present the \FA{} algorithm and
discuss inefficiencies of running it on systolic arrays.
Finally, we motivate the approach of fusing the entire \FA{} on a single systolic array.

\subsection{Matrix Multiplication on Systolic Arrays}\label{sec:systolic-array-intro}

\begin{figure}[t]
\includegraphics[scale=0.75]{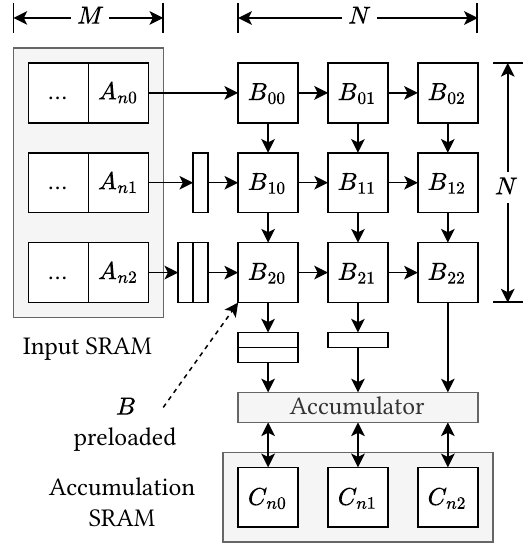}
\caption{Computing $C \mathrel{+}= AB$ on a weight-stationary systolic array.}
\label{fig:systolic-array-example}
\Description{SA}
\end{figure}

Systolic arrays are a class of architectures where data flows through an array of processing elements (PEs) in a rhythmic fashion.
\autoref{fig:systolic-array-example} depicts the basic structure of a systolic array,
comprising $N \times N$ PEs forming a two-dimensional array.
Each PE performs one simple arithmetic operation, such as multiply-accumulate (MAC),
on its input, and drives the output to the next PE, dictated by the array's dataflow.
According to what data is reused in the computation, the dataflow can be categorized as 
\emph{weight-stationary}, \emph{input-stationary}, \emph{output-stationary}, 
or \emph{row-stationary}~\cite{eyeriss-isca}.
As most deep learning workloads favor weight reuse,
many systolic-array-based accelerators,
including Google's TPU~\cite{google-tpu-arch}, adopt the weight-stationary dataflow.

Matrix operations, especially matrix multiplication,
are particularly well-suited for systolic arrays because of their high regularity of computation.
To perform matrix multiplication $C \mathrel{+}= A B$ on a weight-stationary systolic array,
the weight matrix $B$ is preloaded into the array,
while the input matrix $A$, shaped as $M \times N$,
is streamed in from an input SRAM.
To ensure correctness, each row of the input matrix must be delayed by $0$ to $N-1$ cycles
before entering the array. Consequently, each column of the output matrix must be delayed
by $0$ to $N-1$ cycles before use.
The systolic array does not save the output matrix $C$ itself.
Instead, the output matrix is stored in a separate \emph{Accumulation SRAM}
located at the bottom of the array.
The \emph{Accumulator} reads the old values of $C$ from the SRAM,
merges them with the systolic array's output,
and writes the merged values back to the accumulation SRAM.
This near-memory accumulation scheme is widely adopted in both open-source accelerators~\cite{ucb-gemmini}
and commercial products~\cite{trainium-arch}.

During the entire matrix multiplication, MAC computation consumes $M$ cycles,
while weight matrix preloading takes $N$ cycles,
and delaying the input rows and output columns takes another $2 N - 1$ cycles.
In total, the entire computation takes $M + 3N - 1$ cycles to finish,
resulting in a theoretical array utilization of $M / (M + 3N - 1)$. 
When $M \gg N$, the data preparation overhead is negligible.

\subsection{\FA{} on Systolic Arrays}\label{sec:fa-algorithm}

\SetKwInput{kwRequire}{Require}
\tikzHighLight{mat1}{LightSalmon}{{-0.5em, 1em}}{{10em, -0.5em}}
\tikzHighLight{vec}{RoyalBlue}{{-0.5em, 1em}}{{5.7em, -0.5em}}
\tikzHighLight{mat2}{LightSalmon}{{-0.5em, 0.9em}}{{9.05em, -1.7em}}
\begin{algorithm}[t]
    \caption{\FA{}-2 and 3 forward pass}
    \label{alg:flash-attention}
    \kwRequire{Matrices $Q, K, V \in \mathbb{R}^{LEN \times d}$}
    Divide $Q$ into $T_{r} = \lceil \frac{LEN}{B_{r}} \rceil$ blocks of size $B_{r} \times d$ each,
    and divide $K$ and $V$ into $T_{c} = \lceil \frac{LEN}{B_{c}} \rceil$ blocks of size $B_{c} \times d$ each\;
    \For {$ 1 \leq i \leq T_{r} $}
    {
        Initialize $old_m, old_l = (-\infty), (0) \in \mathbb{R}^{B_{r}} $\;
        Initialize $old_O = (0) \in \mathbb{R}^{B_{r} \times d} $\;
        \For {$ 1 \leq j \leq T_{c} $}
        {
            \tikzmark{mat1-start}$ S = Q_{i}K_{j}^{T} \in \mathbb{R}^{B_{r} \times B_{c}} $\tikzmark{mat1-end}\; \label{line:mat1}
            \tikzmark{vec-start}$ local_m = rowmax(S) \in \mathbb{R}^{B_{r}} $\; \label{line:rowmax-start}
            $ new_m = max(local_m, old_m) \in \mathbb{R}^{B_{r}} $\; \label{line:new-m}
            $ a = old_m - new_m \in \mathbb{R}^{B_{r}} $\; \label{line:rowmax-end}
            $ b = exp(\frac{a}{\sqrt{d}}) = exp2(\frac{\log_2 e}{\sqrt{d}}a) \in \mathbb{R}^{B_{r}} $\;
            $ N = S - new_m \in \mathbb{R}^{B_{r} \times B_{c}} $\; \label{line:s-minus-rowmax}
            $ P = exp(\frac{N}{\sqrt{d}}) = exp2(\frac{\log_2 e}{\sqrt{d}}N) \in \mathbb{R}^{B_{r} \times B_{c}} $\; \label{line:exp}
            $ local_l = rowsum(P) \in \mathbb{R}^{B_{r}} $\; \label{line:rowsum}
            $ new_l = old_l \times b + local_l \in \mathbb{R}^{B_{r}} $\tikzmark{vec-end}\; \label{line:l-acc}
            \tikzmark{mat2-start}$ local_O = PV_{j} \in \mathbb{R}^{B_{r} \times d} $\tikzmark{mat2-end}\; \label{line:mat2}
            $ new_O = diag(b)old_O + local_O \in \mathbb{R}^{B_{r} \times d} $\; \label{line:O-acc}
            $ old_m = new_m $\;
            $ old_l = new_l $\;
            $ old_O = new_O $\;
        }
        $ O_{i} = diag(old_l)^{-1}old_O \in \mathbb{R}^{B_{r} \times d} $\; \label{line:re-scaling}
    }
\end{algorithm}

\FA{}~\cite{fa2,fa3} is a memory-efficient attention algorithm that addresses the quadratic memory complexity bottleneck
of standard attention mechanisms.
\FA{} reformulates the attention calculation with tiling and online-softmax to reduce memory usage,
enabling the process of much longer sequences on hardware accelerators with limited on-chip storage.
The \FA{} algorithm is shown in \autoref{alg:flash-attention}.
Same as the official open-source \FA{} implementation~\cite{fa2},
the \texttt{exp} function is implemented as $\exp(x) = 2^{x \log_2 e}$ to benefit from the more
efficient \texttt{exp2} implementation in hardware~\cite{hwexp2}.

\FA{} contains two matrix multiplications (highlighted in red) and various non-matrix-multiplication operations in between,
such as reductions and element-wise computations for softmax (highlighted in blue).
When running \FA{} on systolic-array-based accelerators, the two matrix multiplications are executed on the systolic array,
while other softmax-related operations are offloaded to external vector or scalar units~\cite{trainium-arch}.
This simplistic mapping of computation forces intermediate results to be transferred back and forth between
the systolic array and vector/scalar units through a local SRAM,
essentially destroying any overlap and data reuse between the two matrix multiplications and exaggerating SRAM port contention~\cite{virgo,mosaic}.

To amortize this data movement overhead,
software usually incorporates large tile sizes and schedules multiple Flash\-Attention iterations in a pipelined manner.
However, due to the simplistic mapping of computation with matrix multiplication to the systolic array and softmax to vector/scalar units,
achieving high performance for \FA{} remains challenging~\cite{fusemax},
especially under the silicon constraint that prevents vector/scalar units from providing sufficient \FLOPs{}
to keep up with the systolic array.

\subsection{Fusing \FA{} into Systolic Arrays}

A possible approach to both eliminate the above data movement overhead and improve computation mapping is to discard vector/scalar units and run the entire \FA{} solely on the systolic array.
In this approach, no intermediate results need to be transferred,
and non-matrix-multiplication operations can utilize and share the abundant \FLOPs{} from the systolic array with matrix multiplication,
thereby significantly improving overall hardware utilization.

While seemingly straightforward,
this approach faces the practical challenge of implementing reduction or element-wise operations,
such as \texttt{exp}, \texttt{rowsum}, and \texttt{rowmax}, on the systolic array.
Such implementation is not trivial, as non-linear operations are not a natural fit for systolic arrays.
An efficient implementation without excessively extending the systolic array's PEs or dataflow requires
careful analysis of the arithmetic properties of these operations and novel extensions to the baseline systolic array architecture.
In addition, the numerical accuracy of these operations must be carefully handled to maintain correctness.

Fusing the entire \FA{} into a single systolic array gives us the extra benefit of overlapping executions
of matrix multiplications and softmax operations to further improve hardware utilization.
Similarly, all intermediate results can be generated and consumed in place without resorting to external SRAMs to eliminate SRAM port contention.
Both of these benefits can only be obtained through efficient and fine-grained scheduling of operations in \FA{}.

\section{\SA{} over \MSAGA{}}

Facing the high computational demand of attention kernels, we propose
\MSAGA{}, a novel and hardware-efficient systolic array architecture that
unlocks new capabilities of systolic arrays for \FA{}.
Building on \MSAGA{}, we introduce \SA{}, an optimized kernel that
maximizes hardware utilization by overlapping \FA{} operations within a
single iteration.
In this section, we first present an overview of \MSAGA{} design,
followed by detailed descriptions of how non-matrix-multiplication operations are implemented on \MSAGA{}.
We then elaborate on how \SA{} leverages
\MSAGA{} to efficiently overlap \FA{} operations.

\subsection{\MSAGA{} Design}\label{sec:msaga-design}
\begin{figure}[t]
  \includegraphics[scale=0.75]{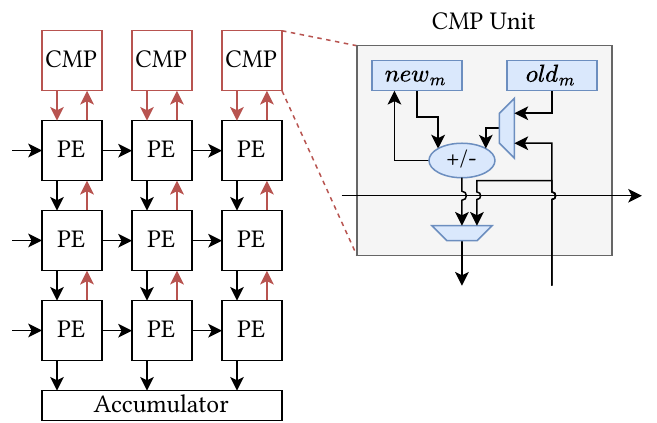}
  \caption{\MSAGA{}'s architectural modifications (highlighted in red) to the standard systolic array.}
  \label{fig:msaga-modifications}
  \Description{\MSAGA{}'s modifications.}
\end{figure}

\MSAGA{} is designed with two main goals:
(1) to reuse the systolic array's existing MAC units and dataflow to minimize hardware overhead;
and (2) to exploit the array's computational pattern to overlap multiple operations.

\autoref{fig:msaga-modifications} shows the architectural modifications to \MSAGA{}.
\MSAGA{} builds on the standard systolic array in \autoref{fig:systolic-array-example}
and introduces the following three key extensions to support non-matrix-multiplication operations in \FA{}.
First, \MSAGA{} introduces an upward data path, enabling column data streaming in both directions.
Second, it adds an array of comparators atop the MAC array to perform column-wise reductions on the fly.
Lastly, each PE is enhanced with the capability for linear interpolation, allowing it to perform non-linear
element-wise operations.

\subsection{Implementing \texttt{rowmax} on \MSAGA{}}

\begin{figure}[t]
  \centering
  \includegraphics[scale=0.75]{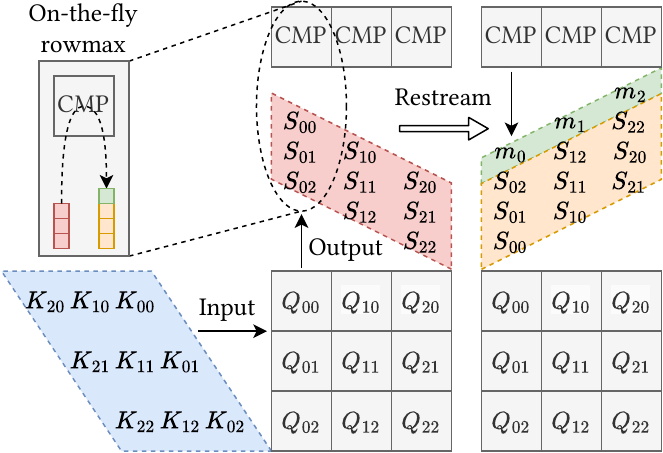}
  \caption{On-the-fly \texttt{rowmax} generation.}
  \label{fig:rowmax}
  \Description{On-the-fly rowmax generation}
\end{figure}

To prepare for \texttt{rowmax}, \MSAGA{} first maps $Q$'s rows to the systolic array's columns, ensuring that rows of $S = QK^T$
(\autoref{line:mat1}, \autoref{alg:flash-attention}) are produced along the columns
of the array.
To compute $local_m = \texttt{rowmax}(S)$
(\autoref{line:rowmax-start}, \autoref{alg:flash-attention}),
\MSAGA{} uses the upward data path to stream each row of $S$ upward to a corresponding
comparator, which performs the maximum operation as a reduction of the entire row on the fly.

As shown in \autoref{fig:msaga-modifications}, each comparator consists of two registers
and a floating-point adder that supports addition, subtraction, and maximum operations.
The $old_m$ register stores the maximum row
value from the previous iteration, while the $new_m$ register keeps track of the
temporary row maximum for the current iteration, updated as each new row value arrives.
The comparator then re-streams each row value back into the systolic array after each $new_m$ update.
By the time the comparator completes the computation
$new_m = \max(local_m, old_m)$
(\autoref{line:new-m}, \autoref{alg:flash-attention}),
the corresponding row of $S$ is resident in the array.
\MSAGA{} then streams $new_m$ (highlighted in green) downward through the systolic
array to compute $N = S - new_m$ directly in place
(\autoref{line:s-minus-rowmax}, \autoref{alg:flash-attention}).

\subsection{Implementing \texttt{exp} on \MSAGA{}}

After \texttt{rowmax} and subtraction operations, an element-wise
exponential function is applied to the matrix $N$, illustrated in
\autoref{line:exp} of \autoref{alg:flash-attention}:
\[
P = exp\left(\frac{N}{\sqrt{d}}\right) =
exp2\left(\frac{\log_2 e}{\sqrt{d}} \cdot N\right).
\]
The constant element-wise multiplication on $N$
can be computed by streaming 
${\log_2 e} / {\sqrt{d}}$ 
from the left of the array as the multiplicand,
as $N$ is already resident in the systolic array 
after the subtraction.

We use \emph{piecewise linear interpolation} (PWL) to approximate the element-wise \texttt{exp2} operation~\cite{pwl}.
Inspired by~\cite{exp_pwl}, which calculates \texttt{exp2} of a fixed-point value, \MSAGA{} applies a similar scheme to decompose
a floating-point number 
$x$ into its integer and fractional components 
$x = x_i + x_f$, where $x_i$ is the integer part and $x_f$ is the fractional part. 
This decomposition leads to the identity:
\[
exp2(x) = 2^{x_i + x_f} = 2^{x_i} \cdot 2^{x_f}.
\]

As $N = S - \texttt{rowmax}(S)$, all elements in $N$ are less than or
equal to zero. Therefore $x_f \in (-1, 0]$, and consequently
$2^{x_f} \in (0.5, 1]$.
Given this limited range for $x_f$, we observe that an 8-piece uniform
PWL is sufficient to keep the relative error
around $10^{-2}$ in the final \FA{} results:
\[
exp2(x) \approx 2^{x_i} \cdot \left(slope_k \cdot x_f + intercept_k \right),
\quad k \in [0, 8),
\]
where the term $2^{x_i}$ only increases the exponent of the final result by $x_i$.

\begin{figure}[t]
  \centering
  \includegraphics[scale=0.75]{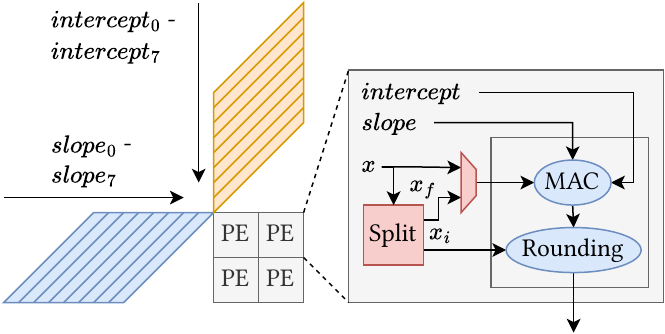}
  \caption{Calculating \texttt{exp2} using PEs extended with PWL capability.}
  \label{fig:exp2-pe}
  \Description{\texttt{exp2} using piecewise linear interpolation.}
\end{figure}

As shown in \autoref{fig:exp2-pe}, the PWL
for $x_f$ can be implemented by reusing the MAC
units in the systolic array. A simple \textit{Split} unit
separates the input $x$ into its integer and fractional parts by shifting the mantissa
bits to align the exponent to zero.
To avoid storing linear interpolation coefficients in the systolic array, we stream $slope_k$
and $intercept_k$ values from the left and top of the array,
respectively.
We observe that all intercepts lie in the range $(0.5, 1]$,
so their exponent can only be $0$ or $-1$. The MSBs of their exponents
can be used to encode the interpolation index $k$.
This enables each PE to update its register based on $k$
without requiring additional control signals.

\subsection{Implementing \texttt{rowsum} on \MSAGA{}}

\begin{figure}[t]
  \centering
  \includegraphics[scale=0.75]{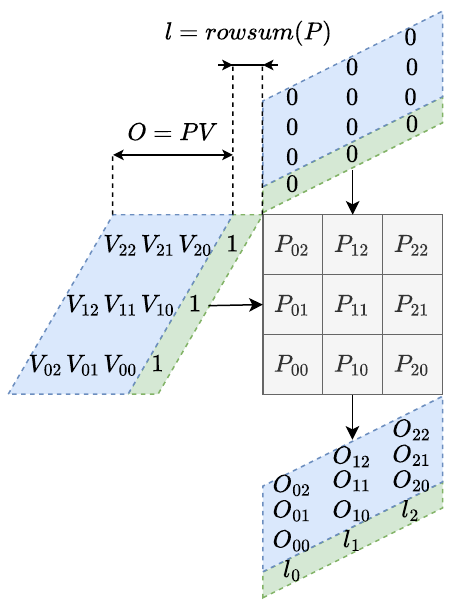}
  \caption{Calculating \texttt{rowsum} using the systolic array.}
  \label{fig:rowsum}
  \Description{Implementation of \texttt{rowsum} using the systolic array.}
\end{figure}

As illustrated in \autoref{fig:rowsum}, the \texttt{rowsum} of $P$ is computed by
streaming a constant one from the left side of the array as the multiplicand,
and streaming zero from the top as the initial addend (highlighted in green).
The second matrix multiplication, $O = PV$
(\autoref{line:mat2}, \autoref{alg:flash-attention}),
can proceed concurrently along the downward data path (highlighted in blue),
starting just one cycle after the \texttt{rowsum} operation begins.
Once the intermediate results $local_l$ and $local_O$ exit the array from the bottom,
they are accumulated into the previous values $old_l$ and $old_O$ in the accumulator.
The updated $new_l$ and $new_O$ are then written back to the accumulation SRAM
(\autoref{line:l-acc} and \autoref{line:O-acc} in \autoref{alg:flash-attention}).

\subsection{Overlapping Computations with \SA{}}\label{sec:overlapping-computations}
\begin{figure*}[t]
  \centering
  \includegraphics[scale=0.75]{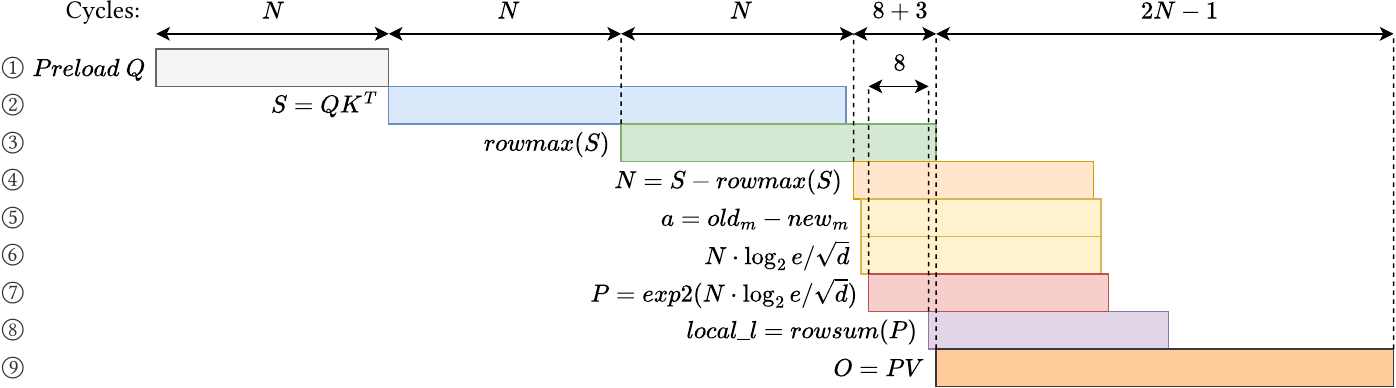}
  \caption{Overlapped computations in \SA{} for the \FA{} inner loop.}
  \label{fig:overlapping}
  \Description{Overlapping computations in the systolic array.}
\end{figure*}

So far, we have introduced the key operations required to perform
\FA{} within a systolic array, including \texttt{rowmax}, \texttt{exp2},
and \texttt{rowsum}.
We now present \SA{}, an optimized kernel that makes the best use of \MSAGA{} to
achieve efficient overlapping of these operations in a single \FA{} iteration.

\autoref{fig:overlapping} illustrates how \SA{} schedules an iteration of \FA{}'s inner loop.
After \circled{1} preloading $Q$ into the systolic array,
\SA{} performs \circled{2} the first matrix multiplication $S = Q K^T$.
It specifically streams the last column of $K$ into the systolic array first (\autoref{fig:rowmax})
to make the first row of $S$ come out first.
Immediately after the first value of $S$ enters the comparator,
\SA{} starts \circled{3} the \texttt{rowmax} calculation, which takes $N$ cycles to complete.
Apart from re-streaming $S$ back to the systolic array,
the comparator also streams $new_m$ downwards,
allowing \circled{4} $N = S - new_m$ to be performed in-place one cycle after $S$ values are made stationary.
Next, \SA{} streams \circled{5} $a = old_m - new_m$ from the comparator,
alongside \circled{6} the constant $\log_2 e / \sqrt{d}$ from the left, propagating $a$ to the accumulator and performing
the constant multiplication $N \cdot \log_2 e / \sqrt{d}$ in parallel immediately
after \circled{4}.
Once this multiplication completes in the top-left PE,
\SA{} initiates \circled{7} the \texttt{exp2} operation,
also starting from the top-left PE. After $k = 8$ cycles of PWL, the result $P = \texttt{exp2}(N \cdot \log_2 e / \sqrt{d})$
is available.
\SA{} then schedules \circled{8} the \texttt{rowsum} of $P$,
starting from the top-left PE, followed by \circled{9} the second matrix
multiplication $O = PV$, both proceeding along the downward data path.

Overall, with \SA{}, \MSAGA{} takes $5 N + 10$ cycles to process an $N \times N$ \FA{} tile.
In contrast, on a naive $N \times N$ systolic array,
the two independent matrix multiplications without any data reuse require up to $8N-2$ cycles
due to preloading and input/output delaying overheads
discussed in \autoref{sec:systolic-array-intro}.

The re-scaling operation in \autoref{line:re-scaling} of \autoref{alg:flash-attention}
is only performed once per outer loop.
It is executed using the accumulator after the second matrix multiplication
completes.
In our implementation, this re-scaling step takes $2N + 20$ cycles,
negligible compared to the inner loop execution time.

\section{\MSAGA{} Microarchitecture}

We implement \MSAGA{} in synthesizable RTL, using Chisel~\cite{chisel} as the hardware description language and Chipyard~\cite{chipyard,chipyard_dac,chipyard_iscas, cook2017diplomatic}
for system-on-chip (SoC) development.
Unlike prior efforts such as Gemmini~\cite{ucb-gemmini} that focus on
design space exploration, our objective is to provide a concrete and complete
implementation to evaluate both the feasibility of the
\SA{} kernel and the associated hardware cost of \MSAGA{}.

\subsection{Microarchitecture Overview}

\begin{figure}[t]
  \includegraphics[scale=0.75]{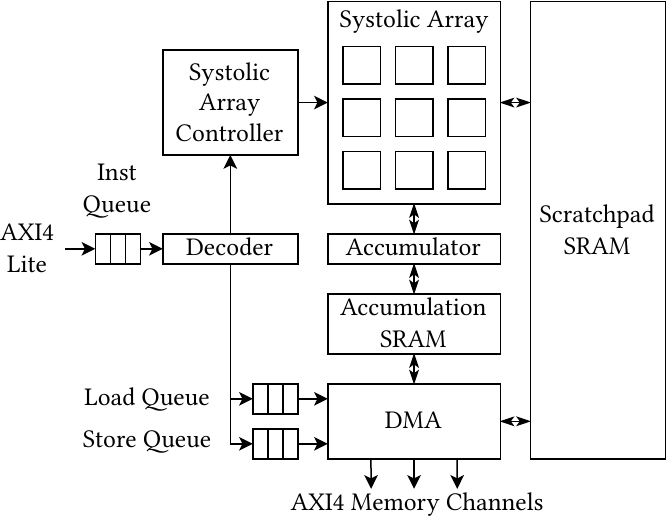}
  \caption{Microarchitecture of the \MSAGA{} accelerator.}
  \label{fig:msaga-microarchitecture}
  \Description{\MSAGA{} microarchitecture overview.}
\end{figure}

\autoref{fig:msaga-microarchitecture} illustrates the microarchitecture of our \MSAGA{} accelerator.
\MSAGA{} receives instructions via an AXI4-Lite interface and accesses backing memory through
a configurable number of AXI4 memory interfaces. Received instructions are buffered in an instruction queue, decoded into
three types: \texttt{load}, \texttt{store}, and \texttt{compute}, and then dispatched
to the load queue, store queue, and systolic array controller, respectively.
Instructions of different types are executed asynchronously, while instructions of the same
type are issued in order with potential interleaving of their execution.

The systolic array controller issues compute instructions once the required data
has been loaded into SRAM. To simplify control logic, \MSAGA{} prioritizes SRAM
access over compute,
so that the latency of compute instructions becomes fully deterministic once issued.
This design allows the controller to statically schedule control signals
based on the progress of each instruction.

The DMA engine follows an iDMA-style interface~\cite{iDMA} and supports 2D data transfers.
Each \texttt{load} or \texttt{store} instruction can specify source and destination
addresses in SRAM and backing memory, along with the transfer size and stride.
When multiple AXI4 channels are available, the DMA engine automatically partitions the transfer
into parallel AXI4 transactions, maximizing memory bandwidth utilization.

\subsection{\MSAGA{} Instruction Set}

To enable fine-grained overlap between various \FA{} operations, the 
systolic array in \MSAGA{} requires more sophisticated control logic 
than standard systolic arrays. The primary design goal of the \MSAGA{} 
instruction set is to ensure that \texttt{compute} instructions execute in a 
fully deterministic manner, independent of memory access latency,
which greatly reduces control complexity.

\begin{figure}[t]
  \includegraphics[scale=0.75]{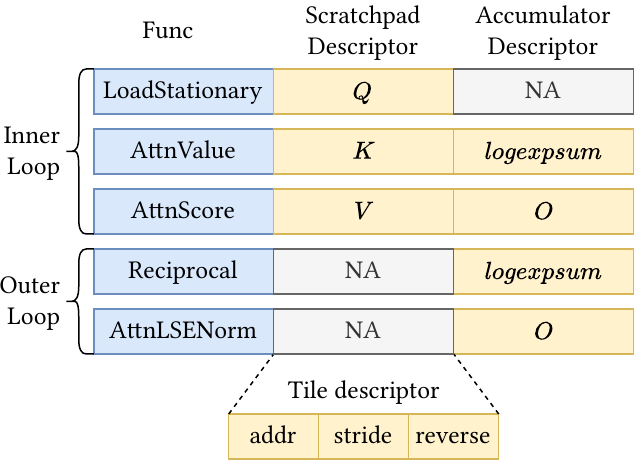}
  \caption{\MSAGA{} compute instructions.}
  \label{fig:msaga-isa}
  \Description{\MSAGA{} compute instructions overview.}
\end{figure}

To this end, we define five compute instructions based on their SRAM 
data dependencies, as illustrated in 
\autoref{fig:msaga-isa}. The \FA{} inner loop is divided into 
three phases:

\begin{itemize}
  \item \texttt{LoadStationary}: preload $Q$ into the systolic array.
  \item \texttt{AttnScore}: perform the first matrix multiplication using 
  $Q$ and $K$, fused with element-wise online softmax to compute the 
  $P$ matrix in-place within the systolic array. The log of the exponent 
  sum for $P$ is also calculated during this step.
  \item \texttt{AttnValue}: perform the second matrix multiplication 
  using $V$ and $P$.
\end{itemize}

The \FA{} outer loop is divided into two additional phases:

\begin{itemize}
  \item \texttt{Reciprocal}: load the log exponent sum of $P$ from the 
  accumulation SRAM and compute its reciprocal. The result is stored 
  in the accumulator as a scaling factor for the next phase.
  \item \texttt{AttnLSENorm}: load the output matrix $O$ from the 
  accumulator and apply the reciprocal scaling factor.
\end{itemize}

Each compute instruction reads one input tile from SRAM and writes one 
output tile to the accumulator SRAM. This one-tile-in, one-tile-out 
design ensures that compute instructions can be executed as soon as 
their corresponding SRAM tile is ready, without waiting for other tiles 
to be loaded or processed.
DMA instructions follow a similar structure to compute instructions, 
consisting of one memory tile descriptor and one SRAM tile descriptor. 
However, they use wider bit fields to support large memory address 
spaces.

\subsection{Systolic Array Controller}

\begin{figure}[t]
  \includegraphics[scale=0.75]{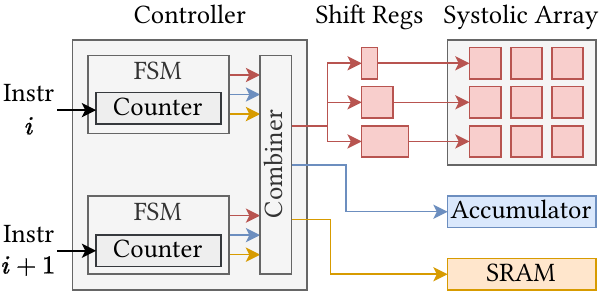}
  \caption{\MSAGA{} systolic array controller.}
  \label{fig:msaga-controller}
  \Description{\MSAGA{} controller overview.}
\end{figure}

The microarchitecture of the systolic array controller is illustrated in \autoref{fig:msaga-controller}.
It mainly comprises two identical finite state machines (FSMs) for generating control signals to the systolic array and
the scratchpad/accumulator SRAM for the current and next \texttt{compute} instruction.
Each FSM contains a counter to track the progress of a \texttt{compute} instruction.
The counter is initialized at the start of the execution, incremented every cycle, and reset to zero upon completion.
As the execution of a \texttt{compute} instruction is fully deterministic,
the FSM simply generates all control signals solely based on the instruction type and the current counter value. 

The dual-FSM architecture of the systolic array controller enables 
efficient operation overlap, as described in \autoref{sec:overlapping-computations}. 
For example, the \texttt{AttnValue} instruction can begin execution 
as soon as the last \texttt{AttnScore} instruction produces the 
first element of the $P$ matrix. The \emph{combiner} unit merges control 
signals from both FSMs, ensuring that all signals are emitted in 
the correct order without conflict.

\begin{figure}[t]
  \centering
  \includegraphics[scale=0.75]{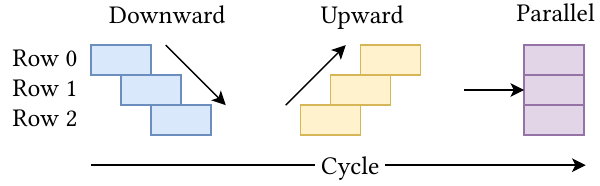}
  \caption{Three PE control flow propagation patterns in the systolic array: 
  Downward, Upward, and Parallel.}
  \label{fig:msaga-dsl}
  \Description{Three PE control flow propagation patterns in the systolic array.}
\end{figure}

There are three control flow propagation patterns in the systolic array: 
\emph{downward}, \emph{upward}, and \emph{parallel}, as illustrated in 
\autoref{fig:msaga-dsl}. The downward pattern propagates control signals 
from the top row to the bottom row; the upward pattern does the reverse; 
and the parallel pattern sends control signals to all rows simultaneously.
The control signals for downward and upward patterns can be generated 
using shift registers without duplicating control logic 
for each row. The parallel pattern can be implemented by broadcasting 
a single control signal to all rows.

To simplify control logic development, we implement a domain-specific 
language (DSL) on top of Chisel and Scala, as shown in \autoref{lst:dsl}. 
This DSL allows designers to specify which control signals should be 
scheduled at each clock cycle. The control logic is then automatically 
synthesized from the DSL specification.

\begin{listing}
\begin{mintedscala}
class PEControl {
  def parallel(start: Int, duration: Int) = ...
  def upward(start: Int, duration: Int) = ...
  def downward(start: Int, duration: Int) = ...
}
class ExampleInstruction(rows: Int, cols: Int)
  extends ExecutionPlan {
  // Load input tile into each row in parallel
  // from cycle 1 to 1 + cols
  load_reg.parallel(1, cols)
  ...
  // Perform MAC operation bottom-up using the
  // upward data path
  mac.upward(1 + cols, rows)
  ...
  // Perform MAC operation top-down using the
  // downward data path
  mac.downward(1 + cols + rows, rows)
}
\end{mintedscala}
\caption{\MSAGA{} Controller DSL.}
\label{lst:dsl}
\end{listing}

\section{\MSAGA{} Kernel Programming Model}

The AWS Neuron Kernel Interface (NKI)~\cite{aws_neuron_nki} provides an expressive 
Python interface that allows users to write custom kernels for AWS Neuron 
devices. It enables instruction-level control with the convenience of 
Python APIs.
Inspired by AWS NKI, we design a similar but simplified Python library 
for writing custom \MSAGA{} kernels. The library consists of three main 
components: type-safe tensors, a Python API for the \MSAGA{} instruction set, 
and a lightweight JIT compiler.

\subsection{Type-Safe Tensors}

The \MSAGA{} Python library implements a PyTorch-like~\cite{pytorch,pytorch2} tensor abstraction 
for representing multi-dimensional arrays. Because the memory hierarchy 
is fully user-managed, tensors can be allocated in one of three memory 
spaces: main memory, scratchpad SRAM, or accumulation SRAM. Accordingly, 
we define three tensor types:

\begin{itemize}
  \item \texttt{MTile} — tensor allocated in main memory,
  \item \texttt{STile} — tensor allocated in scratchpad SRAM,
  \item \texttt{ATile} — tensor allocated in accumulation SRAM.
\end{itemize}

All tensor types support a subset of the PyTorch API, including 
\texttt{shape}, \texttt{dtype}, \texttt{split}, and \texttt{reverse}. 
By explicitly distinguishing tensor types, users can write custom 
kernel functions that clearly declare the expected memory scope of input and output 
tensors.

\subsection{Python API for \MSAGA{} Instructions}

Each \MSAGA{} instruction is exposed through a corresponding Python API. 
These APIs are designed to be type-safe, ensuring correct usage of 
tensor types. The complete set of supported APIs is shown in 
\autoref{lst:msaga_api}.
Internally, these functions construct hardware instructions using the 
metadata of the input tensors, such as shape, stride, and data type. 
This abstraction allows users to write high-level kernels without 
managing low-level hardware details.

\begin{listing}
\begin{mintedpython}
# DMA Instructions
def store_tile(src: ATile, dst: MTile, ...)
def load_tile(src: MTile, dst: STile, ...)
# Compute Instructions
def load_stationary(tile: STile, ...)
def attn_score(K: STile, l: ATile, ...)
def attn_value(V: STile, O: ATile, ...)
def reciprocal(l: ATile, ...)
def attn_lse_norm(O: ATile, ...)
\end{mintedpython}
\caption{\MSAGA{} Python instructions.}
\label{lst:msaga_api}
\end{listing}

\subsection{JIT Kernel Compiler}

The JIT compiler takes a kernel function written with the \MSAGA{} Python 
library and compiles all internally constructed \MSAGA{} instructions into a 
binary program that can be directly submitted to the \MSAGA{} accelerator.
It processes Python functions decorated with \texttt{@F.kernel}, which also
allow users to specify the target device for execution: either a simulated device using 
Verilator~\cite{verilator} or a real device running on Xilinx Alveo U55C FPGA.
Device-host communication is also handled automatically by the compiler.

\autoref{lst:msaga_fa_kernel} shows a complete implementation of the \SA{} kernel using the Python library.
In this example, the device is simulated by Verilator, while the host compiles the entire kernel into a binary program,
loads input tensors into the main memory simulated by DRAMSim2~\cite{dramsim2},
submits the compiled program to the device, and retrieves the output tensors back from DRAMSim2 after the computation is done.
As \MSAGA{} currently does not support matrix transposition,
the host transposes the input matrix $V$ and the output matrix $O$.
On commercial hardware, such transpositions can be performed by the DMA engine 
during memory transfers~\cite{aws_neuron_nki}, and therefore should not incur 
significant overhead.

\begin{listing}
\begin{mintedpython}
import asa as F

@F.kernel(device="verilator_sim")
def attention(Q: MTile, K: MTile, Vt: MTile):
  # allocate output tensor
  Ot: MTile = F.alloc_mem(..)
  # split large tensors into tiles
  Ot_MTiles = Ot.split(br, dim=-1) # [d, br]
  Q_MTiles = Q.split(br, dim=-2)   # [br, d]
  K_MTiles = K.split(bc, dim=-2)   # [br, d]
  Vt_MTiles = Vt.split(bc, dim=-1) # [d, bc]
  # double buffering for Q, K, Vt
  K_STiles = (F.alloc_spad(..),
              F.alloc_spad(..))
  Vt_STiles = (F.alloc_spad(..),
               F.alloc_spad(..))
  # allocate space for accumulation results
  log_expsum = F.alloc_accum(..)   # [1, br]
  Ot_ATile = F.alloc_accum(..)     # [d, br]

  for i, Q_i in enumerate(Q_MTiles):
    F.load_tile(Q_i, Q_STiles[i % 2])
    for j, (K_j, Vt_j) in \
        enumerate(zip(K_MTiles, Vt_MTiles)):
      F.load_stationary(Q_STiles[i % 2])
      F.load_tile(K_j, K_STiles[j % 2])
      F.attn_score(K_STiles[j % 2], log_expsum)
      F.load_tile(Vt_j, Vt_STiles[j % 2])
      F.attn_value(Vt_STiles[j % 2], Ot_ATile)
    F.reciprocal(log_expsum)
    F.attn_lse_norm(Ot_ATile)
    F.store_tile(Ot_ATile, Ot_MTiles[i])

# Run simulation and return result
Ot = attention(Q, K, Vt)
O = Ot.to_numpy().T  # Host-side transpose
\end{mintedpython}
\caption{\MSAGA{} \FA{} kernel.}
\label{lst:msaga_fa_kernel}
\end{listing}



\section{Evaluation}

In this section,
we begin by evaluating \MSAGA{}'s performance compared to state-of-the-art systolic-array-based accelerators.
Then, we assess the accuracy impact of replacing \texttt{exp2} with PWL at the operator, kernel, and model levels.
Finally, we demonstrate \MSAGA{}'s minimal area overhead.

\subsection{\MSAGA{} Performance}\label{sec:eval-performance}

We compare the \FA{} performance of \MSAGA{} against state-of-the-art
systolic-array-based accelerators, 
including Google TPUv5e~\cite{gcp_tpuv5e} and AWS NeuronCore-v2~\cite{neuroncore-v2-arch}.
All accelerators use 16-bit floating-point activation 
and 32-bit accumulation, with the systolic array size of $128 \times 128$. 
\autoref{tab:accelerators-cfg} shows the detailed configurations.

We evaluate the forward pass performance of \FA{} using a single attention head 
on all accelerators. We use \FLOPs{} utilization as the performance metric, 
which reflects the end-to-end absolute performance of each accelerator 
under the same clock frequency, thereby eliminating the impact of 
frequency differences.
\FLOPs{} utilization is defined as:
\[
\text{\FLOPs{} Utilization} = 
\frac{\text{Achieved \FLOPs{}}}{\text{Peak \FLOPs{}}}
\]
where the achieved \FLOPs{} is calculated by:
\[
\text{Achieved \FLOPs{}} = 
\frac{\text{Total FLOPs}}{\text{Execution Time}}
\]
and the total number of FLOPs given by~\cite{fa2}:
\[
\text{Total FLOPs} = 4 \times \text{SeqLen}^2 \times d
\]
where $\text{SeqLen}$ is the sequence length and $d$ is the head dimension.

We use the official \FA{} kernel implementations for the two commercial 
accelerators, which are optimized for their respective architectures:
\begin{itemize}
  \item \textit{jax.experimental.pallas.ops.tpu.flash\_attention} for the TPUv5e~\cite{jax_pallas_tpu}.
  \item \textit{neuronxcc.nki.kernels.attention.flash\_fwd} for the AWS NeuronCore-v2~\cite{aws_neuron_nki}.
\end{itemize}
For \MSAGA{}, we use the \SA{} kernel as shown in \autoref{lst:msaga_fa_kernel}. 
In all experiments, we fix the head dimension at 128 and vary the sequence 
length from 2048 to 16384, without applying causal masking.

\begin{table}[t]
\centering
\begin{threeparttable}
\caption{Hardware configurations of accelerators.}
\label{tab:accelerators-cfg}
\begin{tabular}{cccc}
\toprule
\textbf{Accelerator}               & \textbf{TPUv5e}                          & \textbf{Neuron-v2}  & \textbf{\MSAGA{}}           \\
\midrule
Systolic array size       & $128$                         & $128$        & $128$         \\
Number of arrays & 4                               & 1              & 1               \\
\TFLOPs{} (MAC)     & 196.6                           & 91.75          & 32.77           \\
Frequency                 & 1.5GHz\tnote{1}                         & 2.8GHz         & 1.5GHz            \\
Memory Bandwidth          & 819GB/s                         & 820GB/s        & 820GB/s         \\
Scratchpad SRAM           & \multirow{2}{*}{48MiB} & 24MiB & 192KiB\tnote{2} \\
Accumulation SRAM         &                                 & 2MiB  & 64KiB  \\
Require Vector Unit?      & Yes                    & Yes   & No     \\
\bottomrule
\end{tabular}%
\begin{tablenotes}
  \footnotesize
  \item[1] The clock frequency of TPUv5e is inferred from its reported
  performance using this formula: $\text{freq}_{\text{GHz}} =
  \text{TFLOPS} \times 10^{12} / (2 \times 128 \times 128 \times 10^{9})$.
  \item[2] As \MSAGA{} operates on an $128 \times 128$ tile and only targets \FA{} operations in this experiment, 
  192\,KiB of SRAM is sufficient to support the algorithm.
\end{tablenotes}
\end{threeparttable}
\end{table}

\begin{figure}[t]
  \centering
  \ifreview
  \includegraphics[width=1.0\columnwidth]{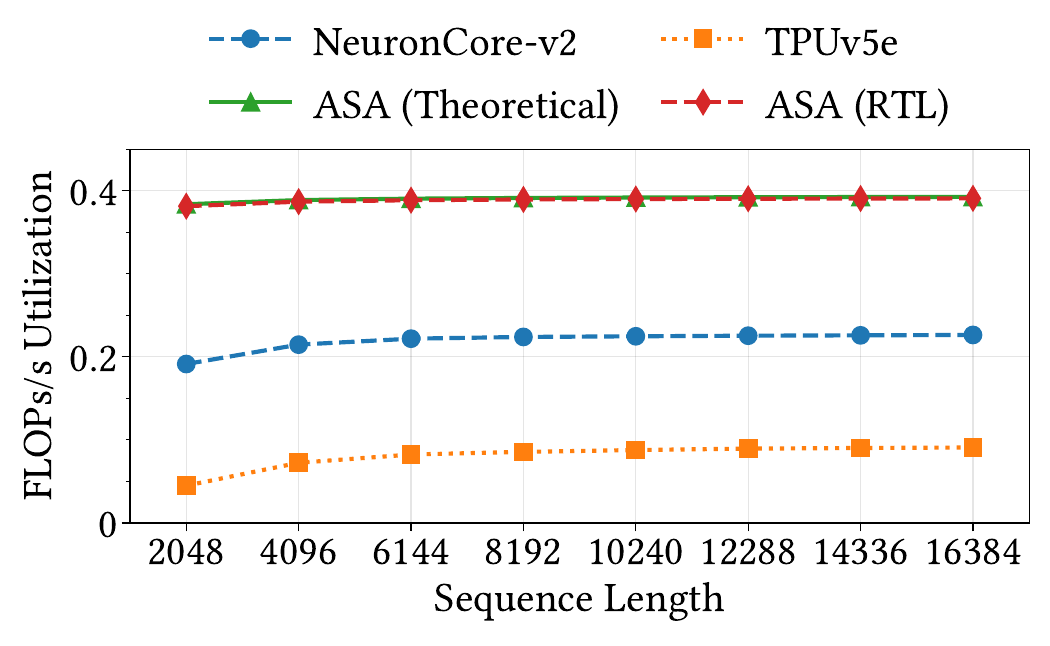}
  \else
  \includegraphics[width=1.0\columnwidth]{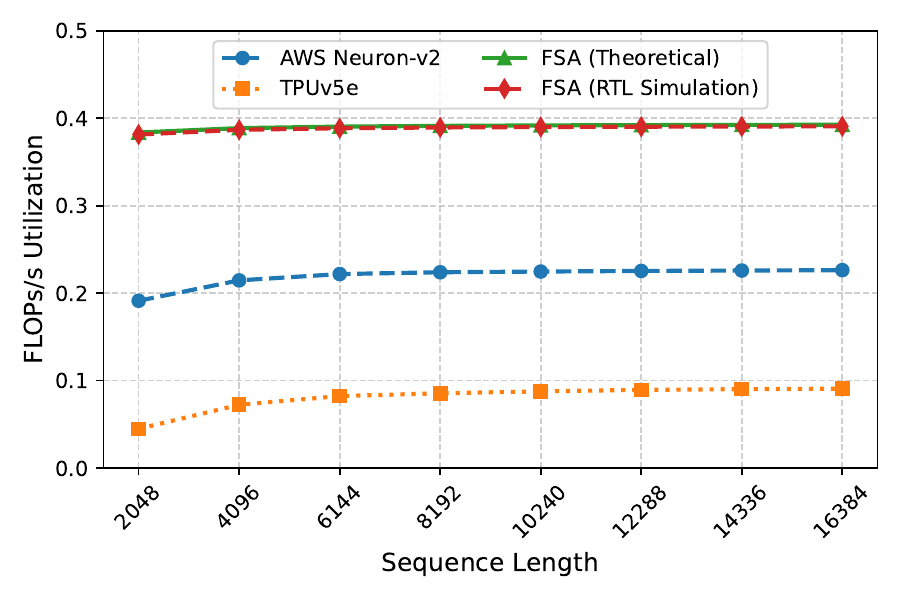}
  \fi
  \caption{\FA{} \FLOPs{} utilization of \MSAGA{} compared to TPUv5e and AWS NeuronCore-v2.}
  \label{fig:perf-cmp}
  \Description{
    The line plot shows the \FLOPs{} utilization of \MSAGA{} compared to TPUv5e and AWS NeuronCore-v2.
  }
\end{figure}

\autoref{fig:perf-cmp} shows the \FLOPs{} utilization of \MSAGA{}, TPUv5e and AWS NeuronCore-v2.
Benefiting from intensive operation overlapping, \MSAGA{} achieves \FLOPs{} utilization 1.77 times and 4.83 times higher
than those of TPUv5e and AWS NeuronCore-v2, respectively.
Moreover, the results confirm that our RTL 
implementation closely aligns with the theoretical performance 
outlined in \autoref{sec:overlapping-computations}, validating the 
correctness of our implementation.

\subsection{\MSAGA{} Accuracy}

The \SA{} produces mathematically equivalent results to standard attention with all orders of floating-point operations preserved. However, as PWL is used for the \texttt{exp2} computation, the final results may 
differ slightly from those generated by commercial accelerators.
In this section, we first analyze the accuracy of PWL 
approximation for the single \texttt{exp2} function. We then assess its overall 
impact on the final attention and model results.

\subsubsection{PWL Accuracy}

\begin{figure}[t]
  \centering
  \includegraphics[width=\columnwidth]{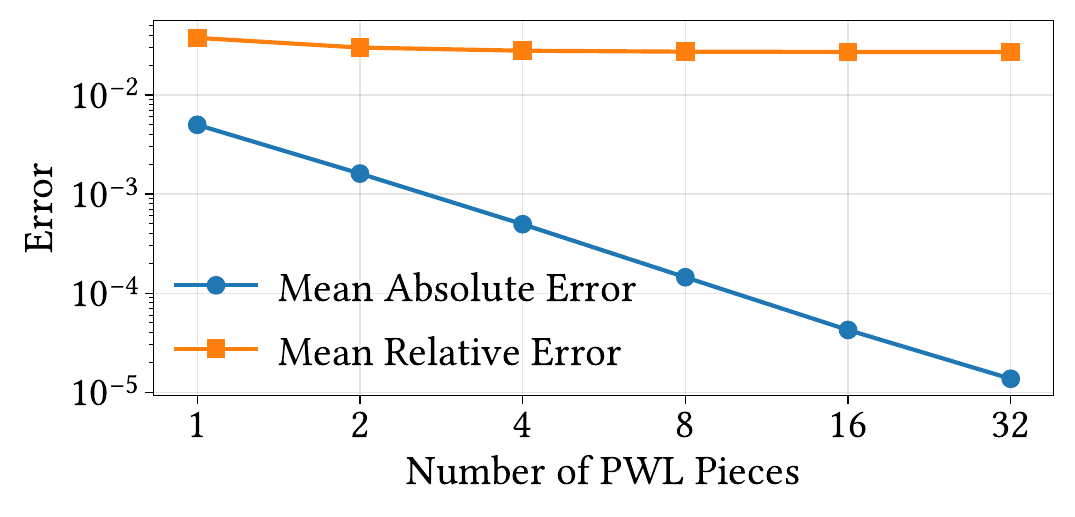}
  \caption{Mean absolute error (MAE) and mean relative error (MRE) of the piecewise linear approximation of \texttt{exp2}.}
  \label{fig:exp2-accuracy}
  \Description{}
\end{figure}

\begin{figure}[t]
  \centering
  \includegraphics[width=\columnwidth]{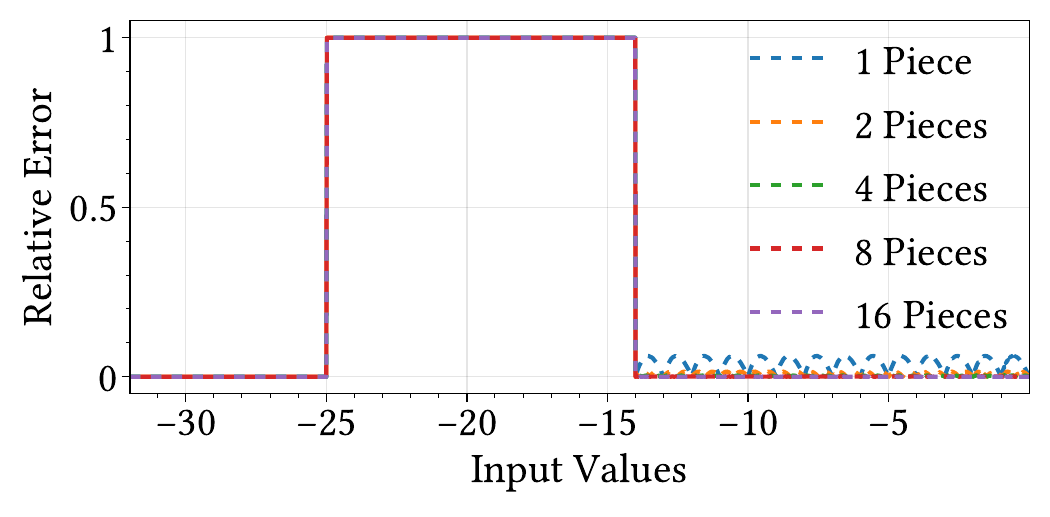}
  \caption{\texttt{exp2} relative error distribution.}
  \label{fig:exp2-rel-distribution}
  \Description{
    The plot shows the mean absolute error (MAE) and mean relative error (MRE) 
    of the piecewise linear approximation of the softmax function as a function 
    of the number of pieces.
  }
\end{figure}

We exhaustively evaluate the PWL approximation over all negative 
\texttt{fp16} values as the input to the \texttt{exp2} function in \FA{} is always negative.
Subnormal values are excluded, as they are typically 
flushed to zero in most accelerators~\cite{googlebfloat16}.

\autoref{fig:exp2-accuracy} shows the mean absolute error (MAE) and 
mean relative error (MRE) of the PWL approximation of 
\texttt{exp2} with various numbers of pieces used. The MAE decreases 
significantly with more segments, while the MRE remains relatively stable. 
This behavior arises because the knots in PWL are 
uniformly spaced, whereas the \texttt{fp16} representation is logarithmic,
making the relative error more sensitive when output values 
are close to zero.

\autoref{fig:exp2-rel-distribution} shows the distribution of relative 
errors across input values. Most relative errors originate from the input 
range $[-25, -14]$, where the output of \texttt{exp2} is close to zero. 
Due to the limited precision of the piecewise approximation, the output 
often rounds to zero, leading to a relative error of 1.0.
For input values less than $-25$, both \texttt{exp2} and PWL output zero,
yielding a relative error of 0.
In our \MSAGA{} implementation, we use 8 segments, which achieves a 
MAE of 0.00014 and an MRE of 0.02728. Although non-uniformly spaced 
knots could reduce the MRE further, we leave this exploration 
to future work.

\subsubsection{\FA{} Accuracy}

We compare results from \SA{} against those from \textit{nn.functional.scaled\_dot\_\-product\_attention} in PyTorch~\cite{pytorch2} to evaluate \FA{} accuracy.
We use the same head dimension and sequence lengths as in the performance evaluation.
Following the same methodology of \FAthree{}~\cite{fa3}, we randomly generate input matrices using the following distribution:
\[
Q, K, V \sim \mathcal{N}(0, 1) + \mathcal{N}(0, 100) \cdot \mathrm{Bernoulli}(0.001).
\]

\begin{table}[t]
\centering
\caption{Mean absolute error (MAE), root mean squared error (RMSE), and mean relative error (MRE) of the \FA{} results on \MSAGA{}.}
\label{table:overall-accuracy}
\begin{tabular}{cccc}
\toprule
\textbf{SeqLen} & \textbf{MAE} & \textbf{RMSE} & \textbf{MRE} \\
\midrule
2048 & 7.983e-03 & 1.315e-02 & 1.558e-02 \\
4096 & 1.379e-02 & 2.290e-02 & 2.596e-02 \\
6144 & 1.849e-02 & 3.085e-02 & 3.545e-02 \\
8192 & 2.253e-02 & 3.772e-02 & 4.413e-02 \\
10240 & 2.595e-02 & 4.373e-02 & 5.259e-02 \\
12288 & 2.890e-02 & 4.873e-02 & 5.920e-02 \\
14336 & 3.165e-02 & 5.351e-02 & 6.529e-02 \\
16384 & 3.403e-02 & 5.784e-02 & 7.181e-02 \\
\bottomrule
\end{tabular}
\end{table}

The overall accuracy results are summarized in \autoref{table:overall-accuracy}.
Across all sequence lengths, MAE lies in the range from $7 \times 10^{-3}$ to $4 \times 10^{-2}$,
while MRE ranges from $1 \times 10^{-2}$ to $8 \times 10^{-2}$,
indicating that approximating \texttt{exp2} with PWL
has a negligible impact on the final attention result.

\subsubsection{End-to-End Model Accuracy}

\begin{table}[t]
\centering
\caption{Perplexity ($\downarrow$ lower is better) comparison between \FA{} 
with \texttt{exp2} and PWL approximations.}
\label{table:perplexity-comparison}
\begin{tabular}{cS[table-format=2.4]S[table-format=2.4]S[table-format=2.4]}
\toprule
\textbf{WikiText2 PPL $\downarrow$} & \textbf{FA-Exp2} & \textbf{FA-PWL} & \textbf{$\Delta$ PPL} \\
\midrule
Llama-3.2-1B & 12.9501 & 12.9492 & -0.0009 \\
Llama-3.2-3B & 10.2997 & 10.2998 & 0.0001 \\
Llama-3.1-8B & 8.1251 & 8.1254 & 0.0002 \\
Gemma-2-2B   & 11.5691 & 11.5680 & -0.0011 \\
Gemma-2-9B   & 9.0789  & 9.0780  & -0.0009 \\
Qwen2.5-0.5B & 19.6371 & 19.6387 & 0.0016 \\
Qwen2.5-1.5B & 13.3067 & 13.3086 & 0.0019 \\
Qwen2.5-3B   & 12.2867 & 12.2860 & -0.0007 \\
Qwen2.5-7B   & 9.5023  & 9.5027  & 0.0005 \\
Qwen2.5-14B  & 6.8394  & 6.8397  & 0.0002 \\
\bottomrule
\end{tabular}
\end{table}

To assess the impact of PWL on the model accuracy, we compare the word perplexity (PPL) of various models, including
Llama~\cite{llama3}, Gemma~\cite{gemma}, and Qwen~\cite{qwen25} families,
using the standard \texttt{exp2} and PWL respectively, with the 
WikiText2 dataset~\cite{wikitext2}.
Since even our largest FPGA cannot accommodate \MSAGA{} with the required $128 \times 128$ systolic array, we model PWL by replacing the \texttt{exp2f} function in the CUDA kernel of 
\FA{} with a software PWL implementation. 
As shown in \autoref{table:perplexity-comparison}, the perplexity with PWL remains almost identical to that of the original 
implementation, indicating that PWL has minimal impact on the performance of realistic LLM models.

\subsection{\MSAGA{} Area}\label{sec:area-overhead}

As described in \autoref{sec:msaga-design}, the area overhead of \MSAGA{} arises from the additional upward data path,
comparators, and Split units.
We synthesize \MSAGA{} systolic array (excluding SRAMs and DMA engines) at 1.5\,GHz using a commercial 16\,nm technology.
The breakdown of the total chip area into various components is shown in \autoref{tab:area}.
The standard systolic array occupies 87.92\% of the total area,
while \MSAGA{}'s additional components only contribute the remaining 12.07\%.
The dominant sources of area overhead come from the upward data path and Split units replicated for every PE,
accounting for 6.24\% and 5.30\% of the total area, respectively.
In contrast, the single array of comparators only consumes 0.53\% of the area.

\begin{table}[t]
\centering
\caption{\MSAGA{} area breakdown.}
\label{tab:area}
\begin{tabular}{@{}llrr@{}}
\toprule
\textbf{Group} & \textbf{Component}              & \textbf{Area (\%)}    & \textbf{Area ($\mu\mathrm{m}^2$)}   \\ \midrule
\multirow{3}{*}{Standard}
               & PEs      & 86.81                 & 24445044                            \\
               & Other logic                     & 1.11                  & 313457                              \\
               & \textit{Total}                  & 87.92                 & 24758501                            \\ \midrule
\multirow{4}{*}{\shortstack{\MSAGA{}\\additional}} 
               & Upward data path                & 6.24                  & 1756641                             \\
               & Split units                     & 5.30                  & 1493150                             \\
               & Comparators                     & 0.53                  & 149524                              \\
               & \textit{Total}                  & 12.07                 & 2029891                             \\ \bottomrule
\end{tabular}
\end{table}

\section{Related Work}

\emph{Systolic Arrays and Spatial Accelerators.}
Extensive work has explored the automatic generation of spatial accelerators for
target applications~\cite{polysa,autosa,dsagen,soda,interstellar,lego}.
These approaches require the application to be written in a specific form and
generate application-specific hardware accelerators.
In contrast, \MSAGA{} is compatible with existing general-purpose systolic arrays
for matrix multiplication.
Stellar~\cite{stellar} and Gemmini~\cite{ucb-gemmini} expose ISAs for
programmability, but they do not support the \FA{} algorithm.
PICACHU~\cite{picachu} proposes a CGRA-based solution to accelerate nonlinear
operations in Transformer models, including the \texttt{exp} in \FA{}.
However, it instantiates a separate array alongside the main MAC systolic array,
increasing area and incurring additional data movement overhead.

\emph{Fusing \FA{} on Systolic Arrays.}
Significant efforts have been made to fuse \FA{} on hardware accelerators. 
COSA~\cite{COSA} and COSA Plus~\cite{COSAPlus} fuse \FA{} into two 
cooperative systolic arrays, where data from the first array must pass 
through a special function unit (SFU) for softmax. Matching the throughput of SFU 
and large systolic arrays can increase hardware cost.  
\FuseMax{}~\cite{fusemax} builds on the cascade Einsum abstraction 
introduced in TeAAL~\cite{teaal} to describe the \FA{} algorithm 
and map it onto spatial accelerator models. 
\FuseMax{} uses a different dataflow than ours: by employing an output-stationary dataflow,
it overlaps four \FA{} 
iterations on the systolic array to obtain high hardware utilization.
However, the overlapping of iterations necessitates storing multiple input tiles in SRAM and
maintaining intermediate results within the array, leading to higher storage overhead 
and control complexity.
Additionally, \FuseMax{} assumes all \FA{} operations are performed in 
\texttt{fp16} format. By contrast, the original \FA{} algorithm uses 
\texttt{fp16} for matrix multiplications but \texttt{fp32} for accumulations. 
Reconciling this difference may require intermediate results to be stored in 
\texttt{fp32}, which can further increase register usage.
ExpMul~\cite{expmul} fuses \texttt{exp} and the second matrix multiplication in \FA{}.
Its approach is orthogonal to ours and could be integrated into \MSAGA{}.

\section{Conclusion}

We present \MSAGA{}, an enhanced systolic array architecture that enables
the entire \FA{} computation within a single systolic array.
Building on \MSAGA{}, we introduce \SA{}, an optimized kernel that efficiently overlaps all \FA{}
operations through careful scheduling and static dataflow management.
The experimental results validate our approach,
demonstrating significantly higher \FLOPs{} utilization than state-of-the-art accelerators,
with negligible accuracy loss and area overhead.

\bibliographystyle{ACM-Reference-Format}
\bibliography{ref}


\begin{thebibliography}{56}


\ifx \showCODEN    \undefined \def \showCODEN     #1{\unskip}     \fi
\ifx \showISBNx    \undefined \def \showISBNx     #1{\unskip}     \fi
\ifx \showISBNxiii \undefined \def \showISBNxiii  #1{\unskip}     \fi
\ifx \showISSN     \undefined \def \showISSN      #1{\unskip}     \fi
\ifx \showLCCN     \undefined \def \showLCCN      #1{\unskip}     \fi
\ifx \shownote     \undefined \def \shownote      #1{#1}          \fi
\ifx \showarticletitle \undefined \def \showarticletitle #1{#1}   \fi
\ifx \showURL      \undefined \def \showURL       {\relax}        \fi
\providecommand\bibfield[2]{#2}
\providecommand\bibinfo[2]{#2}
\providecommand\natexlab[1]{#1}
\providecommand\showeprint[2][]{arXiv:#2}

\bibitem[Agostini et~al\mbox{.}(2022)]%
        {soda}
\bibfield{author}{\bibinfo{person}{Nicolas~Bohm Agostini}, \bibinfo{person}{Ankur Limaye}, \bibinfo{person}{Marco Minutoli}, \bibinfo{person}{Vito~Giovanni Castellana}, \bibinfo{person}{Joseph Manzano}, \bibinfo{person}{Antonino Tumeo}, {and} \bibinfo{person}{Serena Curzel}.} \bibinfo{year}{2022}\natexlab{}.
\newblock \showarticletitle{SODA Synthesizer: an Open-source, Multi-level, Modular, Extensible Compiler from High-level Frameworks to Silicon}. In \bibinfo{booktitle}{\emph{2022 IEEE/ACM International Conference On Computer Aided Design (ICCAD)}}. \bibinfo{pages}{1--7}.
\newblock


\bibitem[Alexandridis et~al\mbox{.}(2025)]%
        {expmul}
\bibfield{author}{\bibinfo{person}{Kosmas Alexandridis}, \bibinfo{person}{Vasileios Titopoulos}, {and} \bibinfo{person}{Giorgos Dimitrakopoulos}.} \bibinfo{year}{2025}\natexlab{}.
\newblock \bibinfo{title}{Low-Cost FlashAttention with Fused Exponential and Multiplication Hardware Operators}.
\newblock
\showeprint[arxiv]{2505.14314}~[cs.AR]
\urldef\tempurl%
\url{https://arxiv.org/abs/2505.14314}
\showURL{%
\tempurl}


\bibitem[Amid et~al\mbox{.}(2020a)]%
        {chipyard_dac}
\bibfield{author}{\bibinfo{person}{Alon Amid}, \bibinfo{person}{David Biancolin}, \bibinfo{person}{Abraham Gonzalez}, \bibinfo{person}{Daniel Grubb}, \bibinfo{person}{Sagar Karandikar}, \bibinfo{person}{Harrison Liew}, \bibinfo{person}{Albert Magyar}, \bibinfo{person}{Howard Mao}, \bibinfo{person}{Albert Ou}, \bibinfo{person}{Nathan Pemberton}, \bibinfo{person}{Paul Rigge}, \bibinfo{person}{Colin Schmidt}, \bibinfo{person}{John Wright}, \bibinfo{person}{Jerry Zhao}, \bibinfo{person}{Jonathan Bachrach}, \bibinfo{person}{Sophia Shao}, \bibinfo{person}{Borivoje Nikolić}, {and} \bibinfo{person}{Krste Asanović}.} \bibinfo{year}{2020}\natexlab{a}.
\newblock \showarticletitle{Invited: Chipyard - An Integrated SoC Research and Implementation Environment}. In \bibinfo{booktitle}{\emph{2020 57th ACM/IEEE Design Automation Conference (DAC)}}. \bibinfo{pages}{1--6}.
\newblock
\href{https://doi.org/10.1109/DAC18072.2020.9218756}{doi:\nolinkurl{10.1109/DAC18072.2020.9218756}}


\bibitem[Amid et~al\mbox{.}(2020b)]%
        {chipyard}
\bibfield{author}{\bibinfo{person}{Alon Amid}, \bibinfo{person}{David Biancolin}, \bibinfo{person}{Abraham Gonzalez}, \bibinfo{person}{Daniel Grubb}, \bibinfo{person}{Sagar Karandikar}, \bibinfo{person}{Harrison Liew}, \bibinfo{person}{Albert Magyar}, \bibinfo{person}{Howard Mao}, \bibinfo{person}{Albert Ou}, \bibinfo{person}{Nathan Pemberton}, \bibinfo{person}{Paul Rigge}, \bibinfo{person}{Colin Schmidt}, \bibinfo{person}{John Wright}, \bibinfo{person}{Jerry Zhao}, \bibinfo{person}{Yakun~Sophia Shao}, \bibinfo{person}{Krste Asanovi\'{c}}, {and} \bibinfo{person}{Borivoje Nikoli\'{c}}.} \bibinfo{year}{2020}\natexlab{b}.
\newblock \showarticletitle{Chipyard: Integrated Design, Simulation, and Implementation Framework for Custom SoCs}.
\newblock \bibinfo{journal}{\emph{IEEE Micro}} \bibinfo{volume}{40}, \bibinfo{number}{4} (\bibinfo{year}{2020}), \bibinfo{pages}{10--21}.
\newblock
\showISSN{1937-4143}
\href{https://doi.org/10.1109/MM.2020.2996616}{doi:\nolinkurl{10.1109/MM.2020.2996616}}


\bibitem[Amid et~al\mbox{.}(2021)]%
        {chipyard_iscas}
\bibfield{author}{\bibinfo{person}{Alon Amid}, \bibinfo{person}{Albert Ou}, \bibinfo{person}{Krste Asanović}, \bibinfo{person}{Yakun~Sophia Shao}, {and} \bibinfo{person}{Borivoje Nikolić}.} \bibinfo{year}{2021}\natexlab{}.
\newblock \showarticletitle{Vertically Integrated Computing Labs Using Open-Source Hardware Generators and Cloud-Hosted FPGAs}. In \bibinfo{booktitle}{\emph{2021 IEEE International Symposium on Circuits and Systems (ISCAS)}}. \bibinfo{pages}{1--5}.
\newblock
\href{https://doi.org/10.1109/ISCAS51556.2021.9401515}{doi:\nolinkurl{10.1109/ISCAS51556.2021.9401515}}


\bibitem[Ansel et~al\mbox{.}(2024)]%
        {pytorch2}
\bibfield{author}{\bibinfo{person}{Jason Ansel}, \bibinfo{person}{Edward Yang}, \bibinfo{person}{Horace He}, \bibinfo{person}{Natalia Gimelshein}, \bibinfo{person}{Animesh Jain}, \bibinfo{person}{Michael Voznesensky}, \bibinfo{person}{Bin Bao}, \bibinfo{person}{Peter Bell}, \bibinfo{person}{David Berard}, \bibinfo{person}{Evgeni Burovski}, \bibinfo{person}{Geeta Chauhan}, \bibinfo{person}{Anjali Chourdia}, \bibinfo{person}{Will Constable}, \bibinfo{person}{Alban Desmaison}, \bibinfo{person}{Zachary DeVito}, \bibinfo{person}{Elias Ellison}, \bibinfo{person}{Will Feng}, \bibinfo{person}{Jiong Gong}, \bibinfo{person}{Michael Gschwind}, \bibinfo{person}{Brian Hirsh}, \bibinfo{person}{Sherlock Huang}, \bibinfo{person}{Kshiteej Kalambarkar}, \bibinfo{person}{Laurent Kirsch}, \bibinfo{person}{Michael Lazos}, \bibinfo{person}{Mario Lezcano}, \bibinfo{person}{Yanbo Liang}, \bibinfo{person}{Jason Liang}, \bibinfo{person}{Yinghai Lu}, \bibinfo{person}{C.~K. Luk}, \bibinfo{person}{Bert Maher}, \bibinfo{person}{Yunjie Pan}, \bibinfo{person}{Christian Puhrsch}, \bibinfo{person}{Matthias Reso}, \bibinfo{person}{Mark Saroufim}, \bibinfo{person}{Marcos~Yukio Siraichi}, \bibinfo{person}{Helen Suk}, \bibinfo{person}{Shunting Zhang}, \bibinfo{person}{Michael Suo}, \bibinfo{person}{Phil Tillet}, \bibinfo{person}{Xu Zhao}, \bibinfo{person}{Eikan Wang}, \bibinfo{person}{Keren Zhou}, \bibinfo{person}{Richard Zou}, \bibinfo{person}{Xiaodong Wang}, \bibinfo{person}{Ajit Mathews}, \bibinfo{person}{William Wen}, \bibinfo{person}{Gregory Chanan}, \bibinfo{person}{Peng Wu}, {and} \bibinfo{person}{Soumith Chintala}.} \bibinfo{year}{2024}\natexlab{}.
\newblock \showarticletitle{PyTorch 2: Faster Machine Learning Through Dynamic Python Bytecode Transformation and Graph Compilation}. In \bibinfo{booktitle}{\emph{Proceedings of the 29th ACM International Conference on Architectural Support for Programming Languages and Operating Systems, Volume 2}} (La Jolla, CA, USA) \emph{(\bibinfo{series}{ASPLOS '24})}. \bibinfo{publisher}{Association for Computing Machinery}, \bibinfo{address}{New York, NY, USA}, \bibinfo{pages}{929–947}.
\newblock
\showISBNx{9798400703850}
\href{https://doi.org/10.1145/3620665.3640366}{doi:\nolinkurl{10.1145/3620665.3640366}}


\bibitem[{AWS Neuron Documentation}(2024a)]%
        {neuroncore-v2-arch}
\bibfield{author}{\bibinfo{person}{{AWS Neuron Documentation}}.} \bibinfo{year}{2024}\natexlab{a}.
\newblock \bibinfo{booktitle}{\emph{NeuronCore-v2 Architecture}}.
\newblock
\urldef\tempurl%
\url{https://awsdocs-neuron.readthedocs-hosted.com/en/latest/general/arch/neuron-hardware/neuron-core-v2.html#neuroncores-v2-arch}
\showURL{%
\tempurl}
\newblock
\shownote{Accessed: 2025-06-15}.


\bibitem[{AWS Neuron Documentation}(2024b)]%
        {trainium-arch}
\bibfield{author}{\bibinfo{person}{{AWS Neuron Documentation}}.} \bibinfo{year}{2024}\natexlab{b}.
\newblock \bibinfo{booktitle}{\emph{Trainium/Inferentia2 Architecture Guide for NKI}}.
\newblock
\urldef\tempurl%
\url{https://awsdocs-neuron.readthedocs-hosted.com/en/latest/general/nki/arch/trainium_inferentia2_arch.html#trainium-inferentia2-arch}
\showURL{%
\tempurl}
\newblock
\shownote{Accessed: 2025-06-15}.


\bibitem[{AWS Neuron Team}(2025)]%
        {aws_neuron_nki}
\bibfield{author}{\bibinfo{person}{{AWS Neuron Team}}.} \bibinfo{year}{2025}\natexlab{}.
\newblock \bibinfo{title}{AWS Neuron NKI Samples}.
\newblock \bibinfo{howpublished}{\url{https://github.com/aws-neuron/nki-samples}}.
\newblock
\newblock
\shownote{Accessed: 2025-06-30}.


\bibitem[Bachrach et~al\mbox{.}(2012)]%
        {chisel}
\bibfield{author}{\bibinfo{person}{Jonathan Bachrach}, \bibinfo{person}{Huy Vo}, \bibinfo{person}{Brian Richards}, \bibinfo{person}{Yunsup Lee}, \bibinfo{person}{Andrew Waterman}, \bibinfo{person}{Rimas Avi\v{z}ienis}, \bibinfo{person}{John Wawrzynek}, {and} \bibinfo{person}{Krste Asanovi\'{c}}.} \bibinfo{year}{2012}\natexlab{}.
\newblock \showarticletitle{Chisel: constructing hardware in a Scala embedded language}. In \bibinfo{booktitle}{\emph{Proceedings of the 49th Annual Design Automation Conference}} (San Francisco, California) \emph{(\bibinfo{series}{DAC '12})}. \bibinfo{publisher}{Association for Computing Machinery}, \bibinfo{address}{New York, NY, USA}, \bibinfo{pages}{1216–1225}.
\newblock
\showISBNx{9781450311991}
\href{https://doi.org/10.1145/2228360.2228584}{doi:\nolinkurl{10.1145/2228360.2228584}}


\bibitem[Benz et~al\mbox{.}(4 01)]%
        {iDMA}
\bibfield{author}{\bibinfo{person}{Thomas Benz}, \bibinfo{person}{Michael Rogenmoser}, \bibinfo{person}{Paul Scheffler}, \bibinfo{person}{Samuel Riedel}, \bibinfo{person}{Alessandro Ottaviano}, \bibinfo{person}{Andreas Kurth}, \bibinfo{person}{Torsten Hoefler}, {and} \bibinfo{person}{Luca Benini}.} \bibinfo{year}{2024-01}\natexlab{}.
\newblock \showarticletitle{A High-performance, Energy-efficient Modular DMA Engine Architecture}.
\newblock \bibinfo{journal}{\emph{IEEE Trans. Comput.}} \bibinfo{volume}{73}, \bibinfo{number}{1} (\bibinfo{year}{2024-01}), \bibinfo{pages}{263 -- 277}.
\newblock
\showISSN{0018-9340}
\href{https://doi.org/10.3929/ethz-b-000641982}{doi:\nolinkurl{10.3929/ethz-b-000641982}}


\bibitem[Brown et~al\mbox{.}(2020)]%
        {gpt3}
\bibfield{author}{\bibinfo{person}{Tom~B. Brown}, \bibinfo{person}{Benjamin Mann}, \bibinfo{person}{Nick Ryder}, \bibinfo{person}{Melanie Subbiah}, \bibinfo{person}{Jared Kaplan}, \bibinfo{person}{Prafulla Dhariwal}, \bibinfo{person}{Arvind Neelakantan}, \bibinfo{person}{Pranav Shyam}, \bibinfo{person}{Girish Sastry}, \bibinfo{person}{Amanda Askell}, \bibinfo{person}{Sandhini Agarwal}, \bibinfo{person}{Ariel Herbert-Voss}, \bibinfo{person}{Gretchen Krueger}, \bibinfo{person}{Tom Henighan}, \bibinfo{person}{Rewon Child}, \bibinfo{person}{Aditya Ramesh}, \bibinfo{person}{Daniel~M. Ziegler}, \bibinfo{person}{Jeffrey Wu}, \bibinfo{person}{Clemens Winter}, \bibinfo{person}{Christopher Hesse}, \bibinfo{person}{Mark Chen}, \bibinfo{person}{Eric Sigler}, \bibinfo{person}{Mateusz Litwin}, \bibinfo{person}{Scott Gray}, \bibinfo{person}{Benjamin Chess}, \bibinfo{person}{Jack Clark}, \bibinfo{person}{Christopher Berner}, \bibinfo{person}{Sam McCandlish}, \bibinfo{person}{Alec Radford}, \bibinfo{person}{Ilya Sutskever}, {and} \bibinfo{person}{Dario Amodei}.} \bibinfo{year}{2020}\natexlab{}.
\newblock \showarticletitle{Language models are few-shot learners}. In \bibinfo{booktitle}{\emph{Proceedings of the 34th International Conference on Neural Information Processing Systems}} (Vancouver, BC, Canada) \emph{(\bibinfo{series}{NIPS '20})}. \bibinfo{publisher}{Curran Associates Inc.}, \bibinfo{address}{Red Hook, NY, USA}, Article \bibinfo{articleno}{159}, \bibinfo{numpages}{25}~pages.
\newblock
\showISBNx{9781713829546}


\bibitem[Carion et~al\mbox{.}(2020)]%
        {transformer-autonomous}
\bibfield{author}{\bibinfo{person}{Nicolas Carion}, \bibinfo{person}{Francisco Massa}, \bibinfo{person}{Gabriel Synnaeve}, \bibinfo{person}{Nicolas Usunier}, \bibinfo{person}{Alexander Kirillov}, {and} \bibinfo{person}{Sergey Zagoruyko}.} \bibinfo{year}{2020}\natexlab{}.
\newblock \showarticletitle{End-to-End Object Detection with Transformers}. In \bibinfo{booktitle}{\emph{Computer Vision – ECCV 2020: 16th European Conference, Glasgow, UK, August 23–28, 2020, Proceedings, Part I}} (Glasgow, United Kingdom). \bibinfo{publisher}{Springer-Verlag}, \bibinfo{address}{Berlin, Heidelberg}, \bibinfo{pages}{213–229}.
\newblock
\showISBNx{978-3-030-58451-1}
\href{https://doi.org/10.1007/978-3-030-58452-8_13}{doi:\nolinkurl{10.1007/978-3-030-58452-8_13}}


\bibitem[Chen et~al\mbox{.}(2016)]%
        {eyeriss-isca}
\bibfield{author}{\bibinfo{person}{Yu-Hsin Chen}, \bibinfo{person}{Joel Emer}, {and} \bibinfo{person}{Vivienne Sze}.} \bibinfo{year}{2016}\natexlab{}.
\newblock \showarticletitle{Eyeriss: A Spatial Architecture for Energy-Efficient Dataflow for Convolutional Neural Networks}. In \bibinfo{booktitle}{\emph{2016 ACM/IEEE 43rd Annual International Symposium on Computer Architecture (ISCA)}}. \bibinfo{pages}{367--379}.
\newblock
\href{https://doi.org/10.1109/ISCA.2016.40}{doi:\nolinkurl{10.1109/ISCA.2016.40}}


\bibitem[Chen et~al\mbox{.}(2017)]%
        {eyeriss-jssc}
\bibfield{author}{\bibinfo{person}{Yu-Hsin Chen}, \bibinfo{person}{Tushar Krishna}, \bibinfo{person}{Joel~S. Emer}, {and} \bibinfo{person}{Vivienne Sze}.} \bibinfo{year}{2017}\natexlab{}.
\newblock \showarticletitle{Eyeriss: An Energy-Efficient Reconfigurable Accelerator for Deep Convolutional Neural Networks}.
\newblock \bibinfo{journal}{\emph{IEEE Journal of Solid-State Circuits}} \bibinfo{volume}{52}, \bibinfo{number}{1} (\bibinfo{year}{2017}), \bibinfo{pages}{127--138}.
\newblock
\href{https://doi.org/10.1109/JSSC.2016.2616357}{doi:\nolinkurl{10.1109/JSSC.2016.2616357}}


\bibitem[Cong and Wang(2018)]%
        {polysa}
\bibfield{author}{\bibinfo{person}{Jason Cong} {and} \bibinfo{person}{Jie Wang}.} \bibinfo{year}{2018}\natexlab{}.
\newblock \showarticletitle{PolySA: Polyhedral-Based Systolic Array Auto-Compilation}. In \bibinfo{booktitle}{\emph{2018 IEEE/ACM International Conference on Computer-Aided Design (ICCAD)}}. \bibinfo{pages}{1--8}.
\newblock
\href{https://doi.org/10.1145/3240765.3240838}{doi:\nolinkurl{10.1145/3240765.3240838}}


\bibitem[Cook et~al\mbox{.}(2017)]%
        {cook2017diplomatic}
\bibfield{author}{\bibinfo{person}{Henry Cook}, \bibinfo{person}{Wesley Terpstra}, {and} \bibinfo{person}{Yunsup Lee}.} \bibinfo{year}{2017}\natexlab{}.
\newblock \showarticletitle{Diplomatic Design Patterns: A TileLink Case Study}. In \bibinfo{booktitle}{\emph{Proceedings of the First Workshop on Computer Architecture Research with RISC--V (CARRV'17)}}. \bibinfo{address}{Boston, MA, USA}, \bibinfo{pages}{1--7}.
\newblock


\bibitem[Dao(2023)]%
        {fa2}
\bibfield{author}{\bibinfo{person}{Tri Dao}.} \bibinfo{year}{2023}\natexlab{}.
\newblock \bibinfo{title}{FlashAttention-2: Faster Attention with Better Parallelism and Work Partitioning}.
\newblock
\showeprint[arxiv]{2307.08691}~[cs.LG]
\urldef\tempurl%
\url{https://arxiv.org/abs/2307.08691}
\showURL{%
\tempurl}


\bibitem[Devlin et~al\mbox{.}(2019)]%
        {bert}
\bibfield{author}{\bibinfo{person}{Jacob Devlin}, \bibinfo{person}{Ming-Wei Chang}, \bibinfo{person}{Kenton Lee}, {and} \bibinfo{person}{Kristina Toutanova}.} \bibinfo{year}{2019}\natexlab{}.
\newblock \showarticletitle{BERT: Pre-training of Deep Bidirectional Transformers for Language Understanding}. In \bibinfo{booktitle}{\emph{North American Chapter of the Association for Computational Linguistics}}.
\newblock
\urldef\tempurl%
\url{https://api.semanticscholar.org/CorpusID:52967399}
\showURL{%
\tempurl}


\bibitem[Genc et~al\mbox{.}(2022)]%
        {ucb-gemmini}
\bibfield{author}{\bibinfo{person}{Hasan Genc}, \bibinfo{person}{Seah Kim}, \bibinfo{person}{Alon Amid}, \bibinfo{person}{Ameer Haj-Ali}, \bibinfo{person}{Vighnesh Iyer}, \bibinfo{person}{Pranav Prakash}, \bibinfo{person}{Jerry Zhao}, \bibinfo{person}{Daniel Grubb}, \bibinfo{person}{Harrison Liew}, \bibinfo{person}{Howard Mao}, \bibinfo{person}{Albert Ou}, \bibinfo{person}{Colin Schmidt}, \bibinfo{person}{Samuel Steffl}, \bibinfo{person}{John Wright}, \bibinfo{person}{Ion Stoica}, \bibinfo{person}{Jonathan Ragan-Kelley}, \bibinfo{person}{Krste Asanovic}, \bibinfo{person}{Borivoje Nikolic}, {and} \bibinfo{person}{Yakun~Sophia Shao}.} \bibinfo{year}{2022}\natexlab{}.
\newblock \bibinfo{booktitle}{\emph{Gemmini: Enabling Systematic Deep-Learning Architecture Evaluation via Full-Stack Integration}}.
\newblock \bibinfo{publisher}{IEEE Press}, \bibinfo{pages}{769–774}.
\newblock
\showISBNx{9781665432740}
\urldef\tempurl%
\url{https://doi.org/10.1109/DAC18074.2021.9586216}
\showURL{%
\tempurl}


\bibitem[Genc et~al\mbox{.}(2024)]%
        {stellar}
\bibfield{author}{\bibinfo{person}{Hasan~Nazim Genc}, \bibinfo{person}{Hansung Kim}, \bibinfo{person}{Prashanth Ganesh}, {and} \bibinfo{person}{Yakun~Sophia Shao}.} \bibinfo{year}{2024}\natexlab{}.
\newblock \showarticletitle{Stellar: An Automated Design Framework for Dense and Sparse Spatial Accelerators}. In \bibinfo{booktitle}{\emph{2024 57th IEEE/ACM International Symposium on Microarchitecture (MICRO)}}. \bibinfo{pages}{409--422}.
\newblock
\href{https://doi.org/10.1109/MICRO61859.2024.00038}{doi:\nolinkurl{10.1109/MICRO61859.2024.00038}}


\bibitem[{Google Cloud}(2019)]%
        {googlebfloat16}
\bibfield{author}{\bibinfo{person}{{Google Cloud}}.} \bibinfo{year}{2019}\natexlab{}.
\newblock \bibinfo{title}{{bfloat16: The secret to high performance on Cloud TPUs}}.
\newblock \bibinfo{howpublished}{\url{https://cloud.google.com/blog/products/ai-machine-learning/bfloat16-the-secret-to-high-performance-on-cloud-tpus}}.
\newblock
\newblock
\shownote{Accessed: 2025-06-27}.


\bibitem[{Google Cloud}(2024)]%
        {google-tpu-arch}
\bibfield{author}{\bibinfo{person}{{Google Cloud}}.} \bibinfo{year}{2024}\natexlab{}.
\newblock \bibinfo{booktitle}{\emph{TPU Architecture}}.
\newblock
\urldef\tempurl%
\url{https://cloud.google.com/tpu/docs/system-architecture-tpu-vm}
\showURL{%
\tempurl}
\newblock
\shownote{Accessed: 2025-06-15}.


\bibitem[{Google Cloud}(2025)]%
        {gcp_tpuv5e}
\bibfield{author}{\bibinfo{person}{{Google Cloud}}.} \bibinfo{year}{2025}\natexlab{}.
\newblock \bibinfo{booktitle}{\emph{{Cloud TPU v5e} Documentation}}.
\newblock
\urldef\tempurl%
\url{https://cloud.google.com/tpu/docs/v5e}
\showURL{%
\tempurl}
\newblock
\shownote{Accessed: 2025-06-15}.


\bibitem[Grattafiori et~al\mbox{.}(2024)]%
        {llama3}
\bibfield{author}{\bibinfo{person}{Aaron Grattafiori}, \bibinfo{person}{Abhimanyu Dubey}, \bibinfo{person}{Abhinav Jauhri}, \bibinfo{person}{Abhinav Pandey}, \bibinfo{person}{Abhishek Kadian}, \bibinfo{person}{Ahmad Al-Dahle}, \bibinfo{person}{Aiesha Letman}, \bibinfo{person}{Akhil Mathur}, \bibinfo{person}{Alan Schelten}, \bibinfo{person}{Alex Vaughan}, \bibinfo{person}{Amy Yang}, \bibinfo{person}{Angela Fan}, \bibinfo{person}{Anirudh Goyal}, \bibinfo{person}{Anthony Hartshorn}, \bibinfo{person}{Aobo Yang}, \bibinfo{person}{Archi Mitra}, \bibinfo{person}{Archie Sravankumar}, \bibinfo{person}{Artem Korenev}, \bibinfo{person}{Arthur Hinsvark}, \bibinfo{person}{Arun Rao}, \bibinfo{person}{Aston Zhang}, \bibinfo{person}{Aurelien Rodriguez}, \bibinfo{person}{Austen Gregerson}, \bibinfo{person}{Ava Spataru}, \bibinfo{person}{Baptiste Roziere}, \bibinfo{person}{Bethany Biron}, \bibinfo{person}{Binh Tang}, \bibinfo{person}{Bobbie Chern}, \bibinfo{person}{Charlotte Caucheteux}, \bibinfo{person}{Chaya Nayak}, \bibinfo{person}{Chloe Bi}, \bibinfo{person}{Chris Marra}, \bibinfo{person}{Chris McConnell}, \bibinfo{person}{Christian Keller}, \bibinfo{person}{Christophe Touret}, \bibinfo{person}{Chunyang Wu}, \bibinfo{person}{Corinne Wong}, \bibinfo{person}{Cristian~Canton Ferrer}, \bibinfo{person}{Cyrus Nikolaidis}, \bibinfo{person}{Damien Allonsius}, \bibinfo{person}{Daniel Song}, \bibinfo{person}{Danielle Pintz}, \bibinfo{person}{Danny Livshits}, \bibinfo{person}{Danny Wyatt}, \bibinfo{person}{David Esiobu}, \bibinfo{person}{Dhruv Choudhary}, \bibinfo{person}{Dhruv Mahajan}, \bibinfo{person}{Diego Garcia-Olano}, \bibinfo{person}{Diego Perino}, \bibinfo{person}{Dieuwke Hupkes}, \bibinfo{person}{Egor Lakomkin}, \bibinfo{person}{Ehab AlBadawy}, \bibinfo{person}{Elina Lobanova}, \bibinfo{person}{Emily Dinan}, \bibinfo{person}{Eric~Michael Smith}, \bibinfo{person}{Filip Radenovic}, \bibinfo{person}{Francisco Guzmán}, \bibinfo{person}{Frank Zhang}, \bibinfo{person}{Gabriel Synnaeve}, \bibinfo{person}{Gabrielle Lee}, \bibinfo{person}{Georgia~Lewis Anderson}, \bibinfo{person}{Govind Thattai}, \bibinfo{person}{Graeme Nail}, \bibinfo{person}{Gregoire Mialon}, \bibinfo{person}{Guan Pang}, \bibinfo{person}{Guillem Cucurell}, \bibinfo{person}{Hailey Nguyen}, \bibinfo{person}{Hannah Korevaar}, \bibinfo{person}{Hu Xu}, \bibinfo{person}{Hugo Touvron}, \bibinfo{person}{Iliyan Zarov}, \bibinfo{person}{Imanol~Arrieta Ibarra}, \bibinfo{person}{Isabel Kloumann}, \bibinfo{person}{Ishan Misra}, \bibinfo{person}{Ivan Evtimov}, \bibinfo{person}{Jack Zhang}, \bibinfo{person}{Jade Copet}, \bibinfo{person}{Jaewon Lee}, \bibinfo{person}{Jan Geffert}, \bibinfo{person}{Jana Vranes}, \bibinfo{person}{Jason Park}, \bibinfo{person}{Jay Mahadeokar}, \bibinfo{person}{Jeet Shah}, \bibinfo{person}{Jelmer van~der Linde}, \bibinfo{person}{Jennifer Billock}, \bibinfo{person}{Jenny Hong}, \bibinfo{person}{Jenya Lee}, \bibinfo{person}{Jeremy Fu}, \bibinfo{person}{Jianfeng Chi}, \bibinfo{person}{Jianyu Huang}, \bibinfo{person}{Jiawen Liu}, \bibinfo{person}{Jie Wang}, \bibinfo{person}{Jiecao Yu}, \bibinfo{person}{Joanna Bitton}, \bibinfo{person}{Joe Spisak}, \bibinfo{person}{Jongsoo Park}, \bibinfo{person}{Joseph Rocca}, \bibinfo{person}{Joshua Johnstun}, \bibinfo{person}{Joshua Saxe}, \bibinfo{person}{Junteng Jia}, \bibinfo{person}{Kalyan~Vasuden Alwala}, \bibinfo{person}{Karthik Prasad}, \bibinfo{person}{Kartikeya Upasani}, \bibinfo{person}{Kate Plawiak}, \bibinfo{person}{Ke Li}, \bibinfo{person}{Kenneth Heafield}, \bibinfo{person}{Kevin Stone}, \bibinfo{person}{Khalid El-Arini}, \bibinfo{person}{Krithika Iyer}, \bibinfo{person}{Kshitiz Malik}, \bibinfo{person}{Kuenley Chiu}, \bibinfo{person}{Kunal Bhalla}, \bibinfo{person}{Kushal Lakhotia}, \bibinfo{person}{Lauren Rantala-Yeary}, \bibinfo{person}{Laurens van~der Maaten}, \bibinfo{person}{Lawrence Chen}, \bibinfo{person}{Liang Tan}, \bibinfo{person}{Liz Jenkins}, \bibinfo{person}{Louis Martin}, \bibinfo{person}{Lovish Madaan}, \bibinfo{person}{Lubo Malo}, \bibinfo{person}{Lukas Blecher}, \bibinfo{person}{Lukas Landzaat}, \bibinfo{person}{Luke de Oliveira}, \bibinfo{person}{Madeline Muzzi}, \bibinfo{person}{Mahesh Pasupuleti}, \bibinfo{person}{Mannat Singh}, \bibinfo{person}{Manohar Paluri}, \bibinfo{person}{Marcin Kardas}, \bibinfo{person}{Maria Tsimpoukelli}, \bibinfo{person}{Mathew Oldham}, \bibinfo{person}{Mathieu Rita}, \bibinfo{person}{Maya Pavlova}, \bibinfo{person}{Melanie Kambadur}, \bibinfo{person}{Mike Lewis}, \bibinfo{person}{Min Si}, \bibinfo{person}{Mitesh~Kumar Singh}, \bibinfo{person}{Mona Hassan}, \bibinfo{person}{Naman Goyal}, \bibinfo{person}{Narjes Torabi}, \bibinfo{person}{Nikolay Bashlykov}, \bibinfo{person}{Nikolay Bogoychev}, \bibinfo{person}{Niladri Chatterji}, \bibinfo{person}{Ning Zhang}, \bibinfo{person}{Olivier Duchenne}, \bibinfo{person}{Onur Çelebi}, \bibinfo{person}{Patrick Alrassy}, \bibinfo{person}{Pengchuan Zhang}, \bibinfo{person}{Pengwei Li}, \bibinfo{person}{Petar Vasic}, \bibinfo{person}{Peter Weng}, \bibinfo{person}{Prajjwal Bhargava}, \bibinfo{person}{Pratik Dubal}, \bibinfo{person}{Praveen Krishnan}, \bibinfo{person}{Punit~Singh Koura}, \bibinfo{person}{Puxin Xu}, \bibinfo{person}{Qing He}, \bibinfo{person}{Qingxiao Dong}, \bibinfo{person}{Ragavan Srinivasan}, \bibinfo{person}{Raj Ganapathy}, \bibinfo{person}{Ramon Calderer}, \bibinfo{person}{Ricardo~Silveira Cabral}, \bibinfo{person}{Robert Stojnic}, \bibinfo{person}{Roberta Raileanu}, \bibinfo{person}{Rohan Maheswari}, \bibinfo{person}{Rohit Girdhar}, \bibinfo{person}{Rohit Patel}, \bibinfo{person}{Romain Sauvestre}, \bibinfo{person}{Ronnie Polidoro}, \bibinfo{person}{Roshan Sumbaly}, \bibinfo{person}{Ross Taylor}, \bibinfo{person}{Ruan Silva}, \bibinfo{person}{Rui Hou}, \bibinfo{person}{Rui Wang}, \bibinfo{person}{Saghar Hosseini}, \bibinfo{person}{Sahana Chennabasappa}, \bibinfo{person}{Sanjay Singh}, \bibinfo{person}{Sean Bell}, \bibinfo{person}{Seohyun~Sonia Kim}, \bibinfo{person}{Sergey Edunov}, \bibinfo{person}{Shaoliang Nie}, \bibinfo{person}{Sharan Narang}, \bibinfo{person}{Sharath Raparthy}, \bibinfo{person}{Sheng Shen}, \bibinfo{person}{Shengye Wan}, \bibinfo{person}{Shruti Bhosale}, \bibinfo{person}{Shun Zhang}, \bibinfo{person}{Simon Vandenhende}, \bibinfo{person}{Soumya Batra}, \bibinfo{person}{Spencer Whitman}, \bibinfo{person}{Sten Sootla}, \bibinfo{person}{Stephane Collot}, \bibinfo{person}{Suchin Gururangan}, \bibinfo{person}{Sydney Borodinsky}, \bibinfo{person}{Tamar Herman}, \bibinfo{person}{Tara Fowler}, \bibinfo{person}{Tarek Sheasha}, \bibinfo{person}{Thomas Georgiou}, \bibinfo{person}{Thomas Scialom}, \bibinfo{person}{Tobias Speckbacher}, \bibinfo{person}{Todor Mihaylov}, \bibinfo{person}{Tong Xiao}, \bibinfo{person}{Ujjwal Karn}, \bibinfo{person}{Vedanuj Goswami}, \bibinfo{person}{Vibhor Gupta}, \bibinfo{person}{Vignesh Ramanathan}, \bibinfo{person}{Viktor Kerkez}, \bibinfo{person}{Vincent Gonguet}, \bibinfo{person}{Virginie Do}, \bibinfo{person}{Vish Vogeti}, \bibinfo{person}{Vítor Albiero}, \bibinfo{person}{Vladan Petrovic}, \bibinfo{person}{Weiwei Chu}, \bibinfo{person}{Wenhan Xiong}, \bibinfo{person}{Wenyin Fu}, \bibinfo{person}{Whitney Meers}, \bibinfo{person}{Xavier Martinet}, \bibinfo{person}{Xiaodong Wang}, \bibinfo{person}{Xiaofang Wang}, \bibinfo{person}{Xiaoqing~Ellen Tan}, \bibinfo{person}{Xide Xia}, \bibinfo{person}{Xinfeng Xie}, \bibinfo{person}{Xuchao Jia}, \bibinfo{person}{Xuewei Wang}, \bibinfo{person}{Yaelle Goldschlag}, \bibinfo{person}{Yashesh Gaur}, \bibinfo{person}{Yasmine Babaei}, \bibinfo{person}{Yi Wen}, \bibinfo{person}{Yiwen Song}, \bibinfo{person}{Yuchen Zhang}, \bibinfo{person}{Yue Li}, \bibinfo{person}{Yuning Mao}, \bibinfo{person}{Zacharie~Delpierre Coudert}, \bibinfo{person}{Zheng Yan}, \bibinfo{person}{Zhengxing Chen}, \bibinfo{person}{Zoe Papakipos}, \bibinfo{person}{Aaditya Singh}, \bibinfo{person}{Aayushi Srivastava}, \bibinfo{person}{Abha Jain}, \bibinfo{person}{Adam Kelsey}, \bibinfo{person}{Adam Shajnfeld}, \bibinfo{person}{Adithya Gangidi}, \bibinfo{person}{Adolfo Victoria}, \bibinfo{person}{Ahuva Goldstand}, \bibinfo{person}{Ajay Menon}, \bibinfo{person}{Ajay Sharma}, \bibinfo{person}{Alex Boesenberg}, \bibinfo{person}{Alexei Baevski}, \bibinfo{person}{Allie Feinstein}, \bibinfo{person}{Amanda Kallet}, \bibinfo{person}{Amit Sangani}, \bibinfo{person}{Amos Teo}, \bibinfo{person}{Anam Yunus}, \bibinfo{person}{Andrei Lupu}, \bibinfo{person}{Andres Alvarado}, \bibinfo{person}{Andrew Caples}, \bibinfo{person}{Andrew Gu}, \bibinfo{person}{Andrew Ho}, \bibinfo{person}{Andrew Poulton}, \bibinfo{person}{Andrew Ryan}, \bibinfo{person}{Ankit Ramchandani}, \bibinfo{person}{Annie Dong}, \bibinfo{person}{Annie Franco}, \bibinfo{person}{Anuj Goyal}, \bibinfo{person}{Aparajita Saraf}, \bibinfo{person}{Arkabandhu Chowdhury}, \bibinfo{person}{Ashley Gabriel}, \bibinfo{person}{Ashwin Bharambe}, \bibinfo{person}{Assaf Eisenman}, \bibinfo{person}{Azadeh Yazdan}, \bibinfo{person}{Beau James}, \bibinfo{person}{Ben Maurer}, \bibinfo{person}{Benjamin Leonhardi}, \bibinfo{person}{Bernie Huang}, \bibinfo{person}{Beth Loyd}, \bibinfo{person}{Beto~De Paola}, \bibinfo{person}{Bhargavi Paranjape}, \bibinfo{person}{Bing Liu}, \bibinfo{person}{Bo Wu}, \bibinfo{person}{Boyu Ni}, \bibinfo{person}{Braden Hancock}, \bibinfo{person}{Bram Wasti}, \bibinfo{person}{Brandon Spence}, \bibinfo{person}{Brani Stojkovic}, \bibinfo{person}{Brian Gamido}, \bibinfo{person}{Britt Montalvo}, \bibinfo{person}{Carl Parker}, \bibinfo{person}{Carly Burton}, \bibinfo{person}{Catalina Mejia}, \bibinfo{person}{Ce Liu}, \bibinfo{person}{Changhan Wang}, \bibinfo{person}{Changkyu Kim}, \bibinfo{person}{Chao Zhou}, \bibinfo{person}{Chester Hu}, \bibinfo{person}{Ching-Hsiang Chu}, \bibinfo{person}{Chris Cai}, \bibinfo{person}{Chris Tindal}, \bibinfo{person}{Christoph Feichtenhofer}, \bibinfo{person}{Cynthia Gao}, \bibinfo{person}{Damon Civin}, \bibinfo{person}{Dana Beaty}, \bibinfo{person}{Daniel Kreymer},
  \bibinfo{person}{Daniel Li}, \bibinfo{person}{David Adkins}, \bibinfo{person}{David Xu}, \bibinfo{person}{Davide Testuggine}, \bibinfo{person}{Delia David}, \bibinfo{person}{Devi Parikh}, \bibinfo{person}{Diana Liskovich}, \bibinfo{person}{Didem Foss}, \bibinfo{person}{Dingkang Wang}, \bibinfo{person}{Duc Le}, \bibinfo{person}{Dustin Holland}, \bibinfo{person}{Edward Dowling}, \bibinfo{person}{Eissa Jamil}, \bibinfo{person}{Elaine Montgomery}, \bibinfo{person}{Eleonora Presani}, \bibinfo{person}{Emily Hahn}, \bibinfo{person}{Emily Wood}, \bibinfo{person}{Eric-Tuan Le}, \bibinfo{person}{Erik Brinkman}, \bibinfo{person}{Esteban Arcaute}, \bibinfo{person}{Evan Dunbar}, \bibinfo{person}{Evan Smothers}, \bibinfo{person}{Fei Sun}, \bibinfo{person}{Felix Kreuk}, \bibinfo{person}{Feng Tian}, \bibinfo{person}{Filippos Kokkinos}, \bibinfo{person}{Firat Ozgenel}, \bibinfo{person}{Francesco Caggioni}, \bibinfo{person}{Frank Kanayet}, \bibinfo{person}{Frank Seide}, \bibinfo{person}{Gabriela~Medina Florez}, \bibinfo{person}{Gabriella Schwarz}, \bibinfo{person}{Gada Badeer}, \bibinfo{person}{Georgia Swee}, \bibinfo{person}{Gil Halpern}, \bibinfo{person}{Grant Herman}, \bibinfo{person}{Grigory Sizov}, \bibinfo{person}{Guangyi}, \bibinfo{person}{Zhang}, \bibinfo{person}{Guna Lakshminarayanan}, \bibinfo{person}{Hakan Inan}, \bibinfo{person}{Hamid Shojanazeri}, \bibinfo{person}{Han Zou}, \bibinfo{person}{Hannah Wang}, \bibinfo{person}{Hanwen Zha}, \bibinfo{person}{Haroun Habeeb}, \bibinfo{person}{Harrison Rudolph}, \bibinfo{person}{Helen Suk}, \bibinfo{person}{Henry Aspegren}, \bibinfo{person}{Hunter Goldman}, \bibinfo{person}{Hongyuan Zhan}, \bibinfo{person}{Ibrahim Damlaj}, \bibinfo{person}{Igor Molybog}, \bibinfo{person}{Igor Tufanov}, \bibinfo{person}{Ilias Leontiadis}, \bibinfo{person}{Irina-Elena Veliche}, \bibinfo{person}{Itai Gat}, \bibinfo{person}{Jake Weissman}, \bibinfo{person}{James Geboski}, \bibinfo{person}{James Kohli}, \bibinfo{person}{Janice Lam}, \bibinfo{person}{Japhet Asher}, \bibinfo{person}{Jean-Baptiste Gaya}, \bibinfo{person}{Jeff Marcus}, \bibinfo{person}{Jeff Tang}, \bibinfo{person}{Jennifer Chan}, \bibinfo{person}{Jenny Zhen}, \bibinfo{person}{Jeremy Reizenstein}, \bibinfo{person}{Jeremy Teboul}, \bibinfo{person}{Jessica Zhong}, \bibinfo{person}{Jian Jin}, \bibinfo{person}{Jingyi Yang}, \bibinfo{person}{Joe Cummings}, \bibinfo{person}{Jon Carvill}, \bibinfo{person}{Jon Shepard}, \bibinfo{person}{Jonathan McPhie}, \bibinfo{person}{Jonathan Torres}, \bibinfo{person}{Josh Ginsburg}, \bibinfo{person}{Junjie Wang}, \bibinfo{person}{Kai Wu}, \bibinfo{person}{Kam~Hou U}, \bibinfo{person}{Karan Saxena}, \bibinfo{person}{Kartikay Khandelwal}, \bibinfo{person}{Katayoun Zand}, \bibinfo{person}{Kathy Matosich}, \bibinfo{person}{Kaushik Veeraraghavan}, \bibinfo{person}{Kelly Michelena}, \bibinfo{person}{Keqian Li}, \bibinfo{person}{Kiran Jagadeesh}, \bibinfo{person}{Kun Huang}, \bibinfo{person}{Kunal Chawla}, \bibinfo{person}{Kyle Huang}, \bibinfo{person}{Lailin Chen}, \bibinfo{person}{Lakshya Garg}, \bibinfo{person}{Lavender A}, \bibinfo{person}{Leandro Silva}, \bibinfo{person}{Lee Bell}, \bibinfo{person}{Lei Zhang}, \bibinfo{person}{Liangpeng Guo}, \bibinfo{person}{Licheng Yu}, \bibinfo{person}{Liron Moshkovich}, \bibinfo{person}{Luca Wehrstedt}, \bibinfo{person}{Madian Khabsa}, \bibinfo{person}{Manav Avalani}, \bibinfo{person}{Manish Bhatt}, \bibinfo{person}{Martynas Mankus}, \bibinfo{person}{Matan Hasson}, \bibinfo{person}{Matthew Lennie}, \bibinfo{person}{Matthias Reso}, \bibinfo{person}{Maxim Groshev}, \bibinfo{person}{Maxim Naumov}, \bibinfo{person}{Maya Lathi}, \bibinfo{person}{Meghan Keneally}, \bibinfo{person}{Miao Liu}, \bibinfo{person}{Michael~L. Seltzer}, \bibinfo{person}{Michal Valko}, \bibinfo{person}{Michelle Restrepo}, \bibinfo{person}{Mihir Patel}, \bibinfo{person}{Mik Vyatskov}, \bibinfo{person}{Mikayel Samvelyan}, \bibinfo{person}{Mike Clark}, \bibinfo{person}{Mike Macey}, \bibinfo{person}{Mike Wang}, \bibinfo{person}{Miquel~Jubert Hermoso}, \bibinfo{person}{Mo Metanat}, \bibinfo{person}{Mohammad Rastegari}, \bibinfo{person}{Munish Bansal}, \bibinfo{person}{Nandhini Santhanam}, \bibinfo{person}{Natascha Parks}, \bibinfo{person}{Natasha White}, \bibinfo{person}{Navyata Bawa}, \bibinfo{person}{Nayan Singhal}, \bibinfo{person}{Nick Egebo}, \bibinfo{person}{Nicolas Usunier}, \bibinfo{person}{Nikhil Mehta}, \bibinfo{person}{Nikolay~Pavlovich Laptev}, \bibinfo{person}{Ning Dong}, \bibinfo{person}{Norman Cheng}, \bibinfo{person}{Oleg Chernoguz}, \bibinfo{person}{Olivia Hart}, \bibinfo{person}{Omkar Salpekar}, \bibinfo{person}{Ozlem Kalinli}, \bibinfo{person}{Parkin Kent}, \bibinfo{person}{Parth Parekh}, \bibinfo{person}{Paul Saab}, \bibinfo{person}{Pavan Balaji}, \bibinfo{person}{Pedro Rittner}, \bibinfo{person}{Philip Bontrager}, \bibinfo{person}{Pierre Roux}, \bibinfo{person}{Piotr Dollar}, \bibinfo{person}{Polina Zvyagina}, \bibinfo{person}{Prashant Ratanchandani}, \bibinfo{person}{Pritish Yuvraj}, \bibinfo{person}{Qian Liang}, \bibinfo{person}{Rachad Alao}, \bibinfo{person}{Rachel Rodriguez}, \bibinfo{person}{Rafi Ayub}, \bibinfo{person}{Raghotham Murthy}, \bibinfo{person}{Raghu Nayani}, \bibinfo{person}{Rahul Mitra}, \bibinfo{person}{Rangaprabhu Parthasarathy}, \bibinfo{person}{Raymond Li}, \bibinfo{person}{Rebekkah Hogan}, \bibinfo{person}{Robin Battey}, \bibinfo{person}{Rocky Wang}, \bibinfo{person}{Russ Howes}, \bibinfo{person}{Ruty Rinott}, \bibinfo{person}{Sachin Mehta}, \bibinfo{person}{Sachin Siby}, \bibinfo{person}{Sai~Jayesh Bondu}, \bibinfo{person}{Samyak Datta}, \bibinfo{person}{Sara Chugh}, \bibinfo{person}{Sara Hunt}, \bibinfo{person}{Sargun Dhillon}, \bibinfo{person}{Sasha Sidorov}, \bibinfo{person}{Satadru Pan}, \bibinfo{person}{Saurabh Mahajan}, \bibinfo{person}{Saurabh Verma}, \bibinfo{person}{Seiji Yamamoto}, \bibinfo{person}{Sharadh Ramaswamy}, \bibinfo{person}{Shaun Lindsay}, \bibinfo{person}{Shaun Lindsay}, \bibinfo{person}{Sheng Feng}, \bibinfo{person}{Shenghao Lin}, \bibinfo{person}{Shengxin~Cindy Zha}, \bibinfo{person}{Shishir Patil}, \bibinfo{person}{Shiva Shankar}, \bibinfo{person}{Shuqiang Zhang}, \bibinfo{person}{Shuqiang Zhang}, \bibinfo{person}{Sinong Wang}, \bibinfo{person}{Sneha Agarwal}, \bibinfo{person}{Soji Sajuyigbe}, \bibinfo{person}{Soumith Chintala}, \bibinfo{person}{Stephanie Max}, \bibinfo{person}{Stephen Chen}, \bibinfo{person}{Steve Kehoe}, \bibinfo{person}{Steve Satterfield}, \bibinfo{person}{Sudarshan Govindaprasad}, \bibinfo{person}{Sumit Gupta}, \bibinfo{person}{Summer Deng}, \bibinfo{person}{Sungmin Cho}, \bibinfo{person}{Sunny Virk}, \bibinfo{person}{Suraj Subramanian}, \bibinfo{person}{Sy Choudhury}, \bibinfo{person}{Sydney Goldman}, \bibinfo{person}{Tal Remez}, \bibinfo{person}{Tamar Glaser}, \bibinfo{person}{Tamara Best}, \bibinfo{person}{Thilo Koehler}, \bibinfo{person}{Thomas Robinson}, \bibinfo{person}{Tianhe Li}, \bibinfo{person}{Tianjun Zhang}, \bibinfo{person}{Tim Matthews}, \bibinfo{person}{Timothy Chou}, \bibinfo{person}{Tzook Shaked}, \bibinfo{person}{Varun Vontimitta}, \bibinfo{person}{Victoria Ajayi}, \bibinfo{person}{Victoria Montanez}, \bibinfo{person}{Vijai Mohan}, \bibinfo{person}{Vinay~Satish Kumar}, \bibinfo{person}{Vishal Mangla}, \bibinfo{person}{Vlad Ionescu}, \bibinfo{person}{Vlad Poenaru}, \bibinfo{person}{Vlad~Tiberiu Mihailescu}, \bibinfo{person}{Vladimir Ivanov}, \bibinfo{person}{Wei Li}, \bibinfo{person}{Wenchen Wang}, \bibinfo{person}{Wenwen Jiang}, \bibinfo{person}{Wes Bouaziz}, \bibinfo{person}{Will Constable}, \bibinfo{person}{Xiaocheng Tang}, \bibinfo{person}{Xiaojian Wu}, \bibinfo{person}{Xiaolan Wang}, \bibinfo{person}{Xilun Wu}, \bibinfo{person}{Xinbo Gao}, \bibinfo{person}{Yaniv Kleinman}, \bibinfo{person}{Yanjun Chen}, \bibinfo{person}{Ye Hu}, \bibinfo{person}{Ye Jia}, \bibinfo{person}{Ye Qi}, \bibinfo{person}{Yenda Li}, \bibinfo{person}{Yilin Zhang}, \bibinfo{person}{Ying Zhang}, \bibinfo{person}{Yossi Adi}, \bibinfo{person}{Youngjin Nam}, \bibinfo{person}{Yu}, \bibinfo{person}{Wang}, \bibinfo{person}{Yu Zhao}, \bibinfo{person}{Yuchen Hao}, \bibinfo{person}{Yundi Qian}, \bibinfo{person}{Yunlu Li}, \bibinfo{person}{Yuzi He}, \bibinfo{person}{Zach Rait}, \bibinfo{person}{Zachary DeVito}, \bibinfo{person}{Zef Rosnbrick}, \bibinfo{person}{Zhaoduo Wen}, \bibinfo{person}{Zhenyu Yang}, \bibinfo{person}{Zhiwei Zhao}, {and} \bibinfo{person}{Zhiyu Ma}.} \bibinfo{year}{2024}\natexlab{}.
\newblock \bibinfo{title}{The Llama 3 Herd of Models}.
\newblock
\showeprint[arxiv]{2407.21783}~[cs.AI]
\urldef\tempurl%
\url{https://arxiv.org/abs/2407.21783}
\showURL{%
\tempurl}


\bibitem[Hussain et~al\mbox{.}(2022)]%
        {hwexp2}
\bibfield{author}{\bibinfo{person}{Muhammad~Awais Hussain}, \bibinfo{person}{Shung-Wei Lin}, {and} \bibinfo{person}{Tsung-Han Tsai}.} \bibinfo{year}{2022}\natexlab{}.
\newblock \showarticletitle{An Area-Efficient and High Throughput Hardware Implementation of Exponent Function}. In \bibinfo{booktitle}{\emph{2022 IEEE International Symposium on Circuits and Systems (ISCAS)}}. \bibinfo{pages}{3369--3372}.
\newblock
\href{https://doi.org/10.1109/ISCAS48785.2022.9937238}{doi:\nolinkurl{10.1109/ISCAS48785.2022.9937238}}


\bibitem[{JAX Developers}({[n.\,d.]})]%
        {jax_pallas_tpu}
\bibfield{author}{\bibinfo{person}{{JAX Developers}}.} \bibinfo{year}{[n.\,d.]}\natexlab{}.
\newblock \bibinfo{title}{JAX Pallas TPU Documentation}.
\newblock \bibinfo{howpublished}{\url{https://docs.jax.dev/en/latest/pallas/tpu/index.html}}.
\newblock
\newblock
\shownote{Accessed: 2025-06-30}.


\bibitem[Jouppi et~al\mbox{.}(2023)]%
        {tpuv4}
\bibfield{author}{\bibinfo{person}{Norm Jouppi}, \bibinfo{person}{George Kurian}, \bibinfo{person}{Sheng Li}, \bibinfo{person}{Peter Ma}, \bibinfo{person}{Rahul Nagarajan}, \bibinfo{person}{Lifeng Nai}, \bibinfo{person}{Nishant Patil}, \bibinfo{person}{Suvinay Subramanian}, \bibinfo{person}{Andy Swing}, \bibinfo{person}{Brian Towles}, \bibinfo{person}{Clifford Young}, \bibinfo{person}{Xiang Zhou}, \bibinfo{person}{Zongwei Zhou}, {and} \bibinfo{person}{David~A Patterson}.} \bibinfo{year}{2023}\natexlab{}.
\newblock \showarticletitle{TPU v4: An Optically Reconfigurable Supercomputer for Machine Learning with Hardware Support for Embeddings}. In \bibinfo{booktitle}{\emph{Proceedings of the 50th Annual International Symposium on Computer Architecture}} (Orlando, FL, USA) \emph{(\bibinfo{series}{ISCA '23})}. \bibinfo{publisher}{Association for Computing Machinery}, \bibinfo{address}{New York, NY, USA}, Article \bibinfo{articleno}{82}, \bibinfo{numpages}{14}~pages.
\newblock
\showISBNx{9798400700958}
\href{https://doi.org/10.1145/3579371.3589350}{doi:\nolinkurl{10.1145/3579371.3589350}}


\bibitem[Jouppi et~al\mbox{.}(2017)]%
        {tpuv1}
\bibfield{author}{\bibinfo{person}{Norman~P. Jouppi}, \bibinfo{person}{Cliff Young}, \bibinfo{person}{Nishant Patil}, \bibinfo{person}{David Patterson}, \bibinfo{person}{Gaurav Agrawal}, \bibinfo{person}{Raminder Bajwa}, \bibinfo{person}{Sarah Bates}, \bibinfo{person}{Suresh Bhatia}, \bibinfo{person}{Nan Boden}, \bibinfo{person}{Al Borchers}, \bibinfo{person}{Rick Boyle}, \bibinfo{person}{Pierre-luc Cantin}, \bibinfo{person}{Clifford Chao}, \bibinfo{person}{Chris Clark}, \bibinfo{person}{Jeremy Coriell}, \bibinfo{person}{Mike Daley}, \bibinfo{person}{Matt Dau}, \bibinfo{person}{Jeffrey Dean}, \bibinfo{person}{Ben Gelb}, \bibinfo{person}{Tara~Vazir Ghaemmaghami}, \bibinfo{person}{Rajendra Gottipati}, \bibinfo{person}{William Gulland}, \bibinfo{person}{Robert Hagmann}, \bibinfo{person}{C.~Richard Ho}, \bibinfo{person}{Doug Hogberg}, \bibinfo{person}{John Hu}, \bibinfo{person}{Robert Hundt}, \bibinfo{person}{Dan Hurt}, \bibinfo{person}{Julian Ibarz}, \bibinfo{person}{Aaron Jaffey}, \bibinfo{person}{Alek Jaworski}, \bibinfo{person}{Alexander Kaplan}, \bibinfo{person}{Harshit Khaitan}, \bibinfo{person}{Daniel Killebrew}, \bibinfo{person}{Andy Koch}, \bibinfo{person}{Naveen Kumar}, \bibinfo{person}{Steve Lacy}, \bibinfo{person}{James Laudon}, \bibinfo{person}{James Law}, \bibinfo{person}{Diemthu Le}, \bibinfo{person}{Chris Leary}, \bibinfo{person}{Zhuyuan Liu}, \bibinfo{person}{Kyle Lucke}, \bibinfo{person}{Alan Lundin}, \bibinfo{person}{Gordon MacKean}, \bibinfo{person}{Adriana Maggiore}, \bibinfo{person}{Maire Mahony}, \bibinfo{person}{Kieran Miller}, \bibinfo{person}{Rahul Nagarajan}, \bibinfo{person}{Ravi Narayanaswami}, \bibinfo{person}{Ray Ni}, \bibinfo{person}{Kathy Nix}, \bibinfo{person}{Thomas Norrie}, \bibinfo{person}{Mark Omernick}, \bibinfo{person}{Narayana Penukonda}, \bibinfo{person}{Andy Phelps}, \bibinfo{person}{Jonathan Ross}, \bibinfo{person}{Matt Ross}, \bibinfo{person}{Amir Salek}, \bibinfo{person}{Emad Samadiani}, \bibinfo{person}{Chris Severn}, \bibinfo{person}{Gregory Sizikov}, \bibinfo{person}{Matthew Snelham}, \bibinfo{person}{Jed Souter}, \bibinfo{person}{Dan Steinberg}, \bibinfo{person}{Andy Swing}, \bibinfo{person}{Mercedes Tan}, \bibinfo{person}{Gregory Thorson}, \bibinfo{person}{Bo Tian}, \bibinfo{person}{Horia Toma}, \bibinfo{person}{Erick Tuttle}, \bibinfo{person}{Vijay Vasudevan}, \bibinfo{person}{Richard Walter}, \bibinfo{person}{Walter Wang}, \bibinfo{person}{Eric Wilcox}, {and} \bibinfo{person}{Doe~Hyun Yoon}.} \bibinfo{year}{2017}\natexlab{}.
\newblock \showarticletitle{In-Datacenter Performance Analysis of a Tensor Processing Unit}.
\newblock \bibinfo{journal}{\emph{SIGARCH Comput. Archit. News}} \bibinfo{volume}{45}, \bibinfo{number}{2} (\bibinfo{date}{June} \bibinfo{year}{2017}), \bibinfo{pages}{1–12}.
\newblock
\showISSN{0163-5964}
\href{https://doi.org/10.1145/3140659.3080246}{doi:\nolinkurl{10.1145/3140659.3080246}}


\bibitem[Kao et~al\mbox{.}(2023)]%
        {flat}
\bibfield{author}{\bibinfo{person}{Sheng-Chun Kao}, \bibinfo{person}{Suvinay Subramanian}, \bibinfo{person}{Gaurav Agrawal}, \bibinfo{person}{Amir Yazdanbakhsh}, {and} \bibinfo{person}{Tushar Krishna}.} \bibinfo{year}{2023}\natexlab{}.
\newblock \showarticletitle{FLAT: An Optimized Dataflow for Mitigating Attention Bottlenecks}. In \bibinfo{booktitle}{\emph{Proceedings of the 28th ACM International Conference on Architectural Support for Programming Languages and Operating Systems, Volume 2}} (Vancouver, BC, Canada) \emph{(\bibinfo{series}{ASPLOS 2023})}. \bibinfo{publisher}{Association for Computing Machinery}, \bibinfo{address}{New York, NY, USA}, \bibinfo{pages}{295–310}.
\newblock
\showISBNx{9781450399166}
\href{https://doi.org/10.1145/3575693.3575747}{doi:\nolinkurl{10.1145/3575693.3575747}}


\bibitem[Kaplan et~al\mbox{.}(2020)]%
        {scaling-laws}
\bibfield{author}{\bibinfo{person}{Jared Kaplan}, \bibinfo{person}{Sam McCandlish}, \bibinfo{person}{Tom Henighan}, \bibinfo{person}{Tom~B. Brown}, \bibinfo{person}{Benjamin Chess}, \bibinfo{person}{Rewon Child}, \bibinfo{person}{Scott Gray}, \bibinfo{person}{Alec Radford}, \bibinfo{person}{Jeffrey Wu}, {and} \bibinfo{person}{Dario Amodei}.} \bibinfo{year}{2020}\natexlab{}.
\newblock \bibinfo{title}{Scaling Laws for Neural Language Models}.
\newblock
\showeprint[arxiv]{2001.08361}~[cs.LG]
\urldef\tempurl%
\url{https://arxiv.org/abs/2001.08361}
\showURL{%
\tempurl}


\bibitem[Kim et~al\mbox{.}(2025)]%
        {virgo}
\bibfield{author}{\bibinfo{person}{Hansung Kim}, \bibinfo{person}{Ruohan~Richard Yan}, \bibinfo{person}{Joshua You}, \bibinfo{person}{Tieliang~Vamber Yang}, {and} \bibinfo{person}{Yakun~Sophia Shao}.} \bibinfo{year}{2025}\natexlab{}.
\newblock \showarticletitle{Virgo: Cluster-level Matrix Unit Integration in GPUs for Scalability and Energy Efficiency}. In \bibinfo{booktitle}{\emph{Proceedings of the 30th ACM International Conference on Architectural Support for Programming Languages and Operating Systems, Volume 2}} (Rotterdam, Netherlands) \emph{(\bibinfo{series}{ASPLOS '25})}. \bibinfo{publisher}{Association for Computing Machinery}, \bibinfo{address}{New York, NY, USA}, \bibinfo{pages}{1382–1399}.
\newblock
\showISBNx{9798400710797}
\href{https://doi.org/10.1145/3676641.3716281}{doi:\nolinkurl{10.1145/3676641.3716281}}


\bibitem[Koca et~al\mbox{.}(2023)]%
        {exp_pwl}
\bibfield{author}{\bibinfo{person}{Nazim~Altar Koca}, \bibinfo{person}{Anh~Tuan Do}, {and} \bibinfo{person}{Chip-Hong Chang}.} \bibinfo{year}{2023}\natexlab{}.
\newblock \showarticletitle{Hardware-efficient Softmax Approximation for Self-Attention Networks}. In \bibinfo{booktitle}{\emph{2023 IEEE International Symposium on Circuits and Systems (ISCAS)}}. \bibinfo{pages}{1--5}.
\newblock
\href{https://doi.org/10.1109/ISCAS46773.2023.10181465}{doi:\nolinkurl{10.1109/ISCAS46773.2023.10181465}}


\bibitem[Kolesnikov et~al\mbox{.}(2021)]%
        {vit}
\bibfield{author}{\bibinfo{person}{Alexander Kolesnikov}, \bibinfo{person}{Alexey Dosovitskiy}, \bibinfo{person}{Dirk Weissenborn}, \bibinfo{person}{Georg Heigold}, \bibinfo{person}{Jakob Uszkoreit}, \bibinfo{person}{Lucas Beyer}, \bibinfo{person}{Matthias Minderer}, \bibinfo{person}{Mostafa Dehghani}, \bibinfo{person}{Neil Houlsby}, \bibinfo{person}{Sylvain Gelly}, \bibinfo{person}{Thomas Unterthiner}, {and} \bibinfo{person}{Xiaohua Zhai}.} \bibinfo{year}{2021}\natexlab{}.
\newblock \showarticletitle{An Image is Worth 16x16 Words: Transformers for Image Recognition at Scale}.
\newblock


\bibitem[Kung(1982)]%
        {original_systolic}
\bibfield{author}{\bibinfo{person}{Kung}.} \bibinfo{year}{1982}\natexlab{}.
\newblock \showarticletitle{Why systolic architectures?}
\newblock \bibinfo{journal}{\emph{Computer}} \bibinfo{volume}{15}, \bibinfo{number}{1} (\bibinfo{year}{1982}), \bibinfo{pages}{37--46}.
\newblock
\href{https://doi.org/10.1109/MC.1982.1653825}{doi:\nolinkurl{10.1109/MC.1982.1653825}}


\bibitem[Lin et~al\mbox{.}(2025)]%
        {lego}
\bibfield{author}{\bibinfo{person}{Yujun Lin}, \bibinfo{person}{Zhekai Zhang}, {and} \bibinfo{person}{Song Han}.} \bibinfo{year}{2025}\natexlab{}.
\newblock \showarticletitle{LEGO: Spatial Accelerator Generation and Optimization for Tensor Applications}. In \bibinfo{booktitle}{\emph{2025 IEEE International Symposium on High Performance Computer Architecture (HPCA)}}. \bibinfo{pages}{1335--1347}.
\newblock
\href{https://doi.org/10.1109/HPCA61900.2025.00101}{doi:\nolinkurl{10.1109/HPCA61900.2025.00101}}


\bibitem[Merity et~al\mbox{.}(2016)]%
        {wikitext2}
\bibfield{author}{\bibinfo{person}{Stephen Merity}, \bibinfo{person}{Caiming Xiong}, \bibinfo{person}{James Bradbury}, {and} \bibinfo{person}{Richard Socher}.} \bibinfo{year}{2016}\natexlab{}.
\newblock \bibinfo{title}{Pointer Sentinel Mixture Models}.
\newblock
\showeprint[arxiv]{1609.07843}~[cs.CL]


\bibitem[Nayak et~al\mbox{.}(2023)]%
        {teaal}
\bibfield{author}{\bibinfo{person}{Nandeeka Nayak}, \bibinfo{person}{Toluwanimi~O. Odemuyiwa}, \bibinfo{person}{Shubham Ugare}, \bibinfo{person}{Christopher Fletcher}, \bibinfo{person}{Michael Pellauer}, {and} \bibinfo{person}{Joel Emer}.} \bibinfo{year}{2023}\natexlab{}.
\newblock \showarticletitle{TeAAL: A Declarative Framework for Modeling Sparse Tensor Accelerators}. In \bibinfo{booktitle}{\emph{Proceedings of the 56th Annual IEEE/ACM International Symposium on Microarchitecture}} (Toronto, ON, Canada) \emph{(\bibinfo{series}{MICRO '23})}. \bibinfo{publisher}{Association for Computing Machinery}, \bibinfo{address}{New York, NY, USA}, \bibinfo{pages}{1255–1270}.
\newblock
\showISBNx{9798400703294}
\href{https://doi.org/10.1145/3613424.3623791}{doi:\nolinkurl{10.1145/3613424.3623791}}


\bibitem[Nayak et~al\mbox{.}(2024)]%
        {fusemax}
\bibfield{author}{\bibinfo{person}{Nandeeka Nayak}, \bibinfo{person}{Xinrui Wu}, \bibinfo{person}{Toluwanimi~O. Odemuyiwa}, \bibinfo{person}{Michael Pellauer}, \bibinfo{person}{Joel~S. Emer}, {and} \bibinfo{person}{Christopher~W. Fletcher}.} \bibinfo{year}{2024}\natexlab{}.
\newblock \showarticletitle{FuseMax: Leveraging Extended Einsums to Optimize Attention Accelerator Design}. In \bibinfo{booktitle}{\emph{2024 57th IEEE/ACM International Symposium on Microarchitecture (MICRO)}}. \bibinfo{pages}{1458--1473}.
\newblock
\href{https://doi.org/10.1109/MICRO61859.2024.00107}{doi:\nolinkurl{10.1109/MICRO61859.2024.00107}}


\bibitem[Ouyang et~al\mbox{.}(2022)]%
        {chatgpt}
\bibfield{author}{\bibinfo{person}{Long Ouyang}, \bibinfo{person}{Jeff Wu}, \bibinfo{person}{Xu Jiang}, \bibinfo{person}{Diogo Almeida}, \bibinfo{person}{Carroll~L. Wainwright}, \bibinfo{person}{Pamela Mishkin}, \bibinfo{person}{Chong Zhang}, \bibinfo{person}{Sandhini Agarwal}, \bibinfo{person}{Katarina Slama}, \bibinfo{person}{Alex Ray}, \bibinfo{person}{John Schulman}, \bibinfo{person}{Jacob Hilton}, \bibinfo{person}{Fraser Kelton}, \bibinfo{person}{Luke Miller}, \bibinfo{person}{Maddie Simens}, \bibinfo{person}{Amanda Askell}, \bibinfo{person}{Peter Welinder}, \bibinfo{person}{Paul Christiano}, \bibinfo{person}{Jan Leike}, {and} \bibinfo{person}{Ryan Lowe}.} \bibinfo{year}{2022}\natexlab{}.
\newblock \bibinfo{title}{Training language models to follow instructions with human feedback}.
\newblock
\showeprint[arxiv]{2203.02155}~[cs.CL]
\urldef\tempurl%
\url{https://arxiv.org/abs/2203.02155}
\showURL{%
\tempurl}


\bibitem[Paszke et~al\mbox{.}(2019)]%
        {pytorch}
\bibfield{author}{\bibinfo{person}{Adam Paszke}, \bibinfo{person}{Sam Gross}, \bibinfo{person}{Francisco Massa}, \bibinfo{person}{Adam Lerer}, \bibinfo{person}{James Bradbury}, \bibinfo{person}{Gregory Chanan}, \bibinfo{person}{Trevor Killeen}, \bibinfo{person}{Zeming Lin}, \bibinfo{person}{Natalia Gimelshein}, \bibinfo{person}{Luca Antiga}, \bibinfo{person}{Alban Desmaison}, \bibinfo{person}{Andreas K\"{o}pf}, \bibinfo{person}{Edward Yang}, \bibinfo{person}{Zach DeVito}, \bibinfo{person}{Martin Raison}, \bibinfo{person}{Alykhan Tejani}, \bibinfo{person}{Sasank Chilamkurthy}, \bibinfo{person}{Benoit Steiner}, \bibinfo{person}{Lu Fang}, \bibinfo{person}{Junjie Bai}, {and} \bibinfo{person}{Soumith Chintala}.} \bibinfo{year}{2019}\natexlab{}.
\newblock \bibinfo{booktitle}{\emph{PyTorch: an imperative style, high-performance deep learning library}}.
\newblock \bibinfo{publisher}{Curran Associates Inc.}, \bibinfo{address}{Red Hook, NY, USA}.
\newblock


\bibitem[Qin et~al\mbox{.}(2025)]%
        {picachu}
\bibfield{author}{\bibinfo{person}{Jiajun Qin}, \bibinfo{person}{Tianhua Xia}, \bibinfo{person}{Cheng Tan}, \bibinfo{person}{Jeff Zhang}, {and} \bibinfo{person}{Sai~Qian Zhang}.} \bibinfo{year}{2025}\natexlab{}.
\newblock \showarticletitle{PICACHU: Plug-In CGRA Handling Upcoming Nonlinear Operations in LLMs}. In \bibinfo{booktitle}{\emph{Proceedings of the 30th ACM International Conference on Architectural Support for Programming Languages and Operating Systems, Volume 2}} (Rotterdam, Netherlands) \emph{(\bibinfo{series}{ASPLOS '25})}. \bibinfo{publisher}{Association for Computing Machinery}, \bibinfo{address}{New York, NY, USA}, \bibinfo{pages}{845–861}.
\newblock
\showISBNx{9798400710797}
\href{https://doi.org/10.1145/3676641.3716013}{doi:\nolinkurl{10.1145/3676641.3716013}}


\bibitem[Rosenfeld et~al\mbox{.}(2011)]%
        {dramsim2}
\bibfield{author}{\bibinfo{person}{Paul Rosenfeld}, \bibinfo{person}{Elliott Cooper-Balis}, {and} \bibinfo{person}{Bruce Jacob}.} \bibinfo{year}{2011}\natexlab{}.
\newblock \showarticletitle{DRAMSim2: A Cycle Accurate Memory System Simulator}.
\newblock \bibinfo{journal}{\emph{IEEE Computer Architecture Letters}} \bibinfo{volume}{10}, \bibinfo{number}{1} (\bibinfo{year}{2011}), \bibinfo{pages}{16--19}.
\newblock
\href{https://doi.org/10.1109/L-CA.2011.4}{doi:\nolinkurl{10.1109/L-CA.2011.4}}


\bibitem[Shah et~al\mbox{.}(2024)]%
        {fa3}
\bibfield{author}{\bibinfo{person}{Jay Shah}, \bibinfo{person}{Ganesh Bikshandi}, \bibinfo{person}{Ying Zhang}, \bibinfo{person}{Vijay Thakkar}, \bibinfo{person}{Pradeep Ramani}, {and} \bibinfo{person}{Tri Dao}.} \bibinfo{year}{2024}\natexlab{}.
\newblock \bibinfo{title}{FlashAttention-3: Fast and Accurate Attention with Asynchrony and Low-precision}.
\newblock
\showeprint[arxiv]{2407.08608}~[cs.LG]
\urldef\tempurl%
\url{https://arxiv.org/abs/2407.08608}
\showURL{%
\tempurl}


\bibitem[Strollo et~al\mbox{.}(2011)]%
        {pwl}
\bibfield{author}{\bibinfo{person}{Antonio~G.M. Strollo}, \bibinfo{person}{Davide De~Caro}, {and} \bibinfo{person}{Nicola Petra}.} \bibinfo{year}{2011}\natexlab{}.
\newblock \showarticletitle{Elementary Functions Hardware Implementation Using Constrained Piecewise-Polynomial Approximations}.
\newblock \bibinfo{journal}{\emph{IEEE Trans. Comput.}} \bibinfo{volume}{60}, \bibinfo{number}{3} (\bibinfo{year}{2011}), \bibinfo{pages}{418--432}.
\newblock
\href{https://doi.org/10.1109/TC.2010.127}{doi:\nolinkurl{10.1109/TC.2010.127}}


\bibitem[Strubell et~al\mbox{.}(2019)]%
        {energy-scaling-laws}
\bibfield{author}{\bibinfo{person}{Emma Strubell}, \bibinfo{person}{Ananya Ganesh}, {and} \bibinfo{person}{Andrew McCallum}.} \bibinfo{year}{2019}\natexlab{}.
\newblock \showarticletitle{Energy and Policy Considerations for Deep Learning in NLP}.
\newblock \bibinfo{journal}{\emph{ArXiv}}  \bibinfo{volume}{abs/1906.02243} (\bibinfo{year}{2019}).
\newblock
\urldef\tempurl%
\url{https://api.semanticscholar.org/CorpusID:174802812}
\showURL{%
\tempurl}


\bibitem[Team et~al\mbox{.}(2024)]%
        {gemma}
\bibfield{author}{\bibinfo{person}{Gemma Team}, \bibinfo{person}{Morgane Riviere}, \bibinfo{person}{Shreya Pathak}, \bibinfo{person}{Pier~Giuseppe Sessa}, \bibinfo{person}{Cassidy Hardin}, \bibinfo{person}{Surya Bhupatiraju}, \bibinfo{person}{Léonard Hussenot}, \bibinfo{person}{Thomas Mesnard}, \bibinfo{person}{Bobak Shahriari}, \bibinfo{person}{Alexandre Ramé}, \bibinfo{person}{Johan Ferret}, \bibinfo{person}{Peter Liu}, \bibinfo{person}{Pouya Tafti}, \bibinfo{person}{Abe Friesen}, \bibinfo{person}{Michelle Casbon}, \bibinfo{person}{Sabela Ramos}, \bibinfo{person}{Ravin Kumar}, \bibinfo{person}{Charline~Le Lan}, \bibinfo{person}{Sammy Jerome}, \bibinfo{person}{Anton Tsitsulin}, \bibinfo{person}{Nino Vieillard}, \bibinfo{person}{Piotr Stanczyk}, \bibinfo{person}{Sertan Girgin}, \bibinfo{person}{Nikola Momchev}, \bibinfo{person}{Matt Hoffman}, \bibinfo{person}{Shantanu Thakoor}, \bibinfo{person}{Jean-Bastien Grill}, \bibinfo{person}{Behnam Neyshabur}, \bibinfo{person}{Olivier Bachem}, \bibinfo{person}{Alanna Walton}, \bibinfo{person}{Aliaksei Severyn}, \bibinfo{person}{Alicia Parrish}, \bibinfo{person}{Aliya Ahmad}, \bibinfo{person}{Allen Hutchison}, \bibinfo{person}{Alvin Abdagic}, \bibinfo{person}{Amanda Carl}, \bibinfo{person}{Amy Shen}, \bibinfo{person}{Andy Brock}, \bibinfo{person}{Andy Coenen}, \bibinfo{person}{Anthony Laforge}, \bibinfo{person}{Antonia Paterson}, \bibinfo{person}{Ben Bastian}, \bibinfo{person}{Bilal Piot}, \bibinfo{person}{Bo Wu}, \bibinfo{person}{Brandon Royal}, \bibinfo{person}{Charlie Chen}, \bibinfo{person}{Chintu Kumar}, \bibinfo{person}{Chris Perry}, \bibinfo{person}{Chris Welty}, \bibinfo{person}{Christopher~A. Choquette-Choo}, \bibinfo{person}{Danila Sinopalnikov}, \bibinfo{person}{David Weinberger}, \bibinfo{person}{Dimple Vijaykumar}, \bibinfo{person}{Dominika Rogozińska}, \bibinfo{person}{Dustin Herbison}, \bibinfo{person}{Elisa Bandy}, \bibinfo{person}{Emma Wang}, \bibinfo{person}{Eric Noland}, \bibinfo{person}{Erica Moreira}, \bibinfo{person}{Evan Senter}, \bibinfo{person}{Evgenii Eltyshev}, \bibinfo{person}{Francesco Visin}, \bibinfo{person}{Gabriel Rasskin}, \bibinfo{person}{Gary Wei}, \bibinfo{person}{Glenn Cameron}, \bibinfo{person}{Gus Martins}, \bibinfo{person}{Hadi Hashemi}, \bibinfo{person}{Hanna Klimczak-Plucińska}, \bibinfo{person}{Harleen Batra}, \bibinfo{person}{Harsh Dhand}, \bibinfo{person}{Ivan Nardini}, \bibinfo{person}{Jacinda Mein}, \bibinfo{person}{Jack Zhou}, \bibinfo{person}{James Svensson}, \bibinfo{person}{Jeff Stanway}, \bibinfo{person}{Jetha Chan}, \bibinfo{person}{Jin~Peng Zhou}, \bibinfo{person}{Joana Carrasqueira}, \bibinfo{person}{Joana Iljazi}, \bibinfo{person}{Jocelyn Becker}, \bibinfo{person}{Joe Fernandez}, \bibinfo{person}{Joost van Amersfoort}, \bibinfo{person}{Josh Gordon}, \bibinfo{person}{Josh Lipschultz}, \bibinfo{person}{Josh Newlan}, \bibinfo{person}{Ju yeong Ji}, \bibinfo{person}{Kareem Mohamed}, \bibinfo{person}{Kartikeya Badola}, \bibinfo{person}{Kat Black}, \bibinfo{person}{Katie Millican}, \bibinfo{person}{Keelin McDonell}, \bibinfo{person}{Kelvin Nguyen}, \bibinfo{person}{Kiranbir Sodhia}, \bibinfo{person}{Kish Greene}, \bibinfo{person}{Lars~Lowe Sjoesund}, \bibinfo{person}{Lauren Usui}, \bibinfo{person}{Laurent Sifre}, \bibinfo{person}{Lena Heuermann}, \bibinfo{person}{Leticia Lago}, \bibinfo{person}{Lilly McNealus}, \bibinfo{person}{Livio~Baldini Soares}, \bibinfo{person}{Logan Kilpatrick}, \bibinfo{person}{Lucas Dixon}, \bibinfo{person}{Luciano Martins}, \bibinfo{person}{Machel Reid}, \bibinfo{person}{Manvinder Singh}, \bibinfo{person}{Mark Iverson}, \bibinfo{person}{Martin Görner}, \bibinfo{person}{Mat Velloso}, \bibinfo{person}{Mateo Wirth}, \bibinfo{person}{Matt Davidow}, \bibinfo{person}{Matt Miller}, \bibinfo{person}{Matthew Rahtz}, \bibinfo{person}{Matthew Watson}, \bibinfo{person}{Meg Risdal}, \bibinfo{person}{Mehran Kazemi}, \bibinfo{person}{Michael Moynihan}, \bibinfo{person}{Ming Zhang}, \bibinfo{person}{Minsuk Kahng}, \bibinfo{person}{Minwoo Park}, \bibinfo{person}{Mofi Rahman}, \bibinfo{person}{Mohit Khatwani}, \bibinfo{person}{Natalie Dao}, \bibinfo{person}{Nenshad Bardoliwalla}, \bibinfo{person}{Nesh Devanathan}, \bibinfo{person}{Neta Dumai}, \bibinfo{person}{Nilay Chauhan}, \bibinfo{person}{Oscar Wahltinez}, \bibinfo{person}{Pankil Botarda}, \bibinfo{person}{Parker Barnes}, \bibinfo{person}{Paul Barham}, \bibinfo{person}{Paul Michel}, \bibinfo{person}{Pengchong Jin}, \bibinfo{person}{Petko Georgiev}, \bibinfo{person}{Phil Culliton}, \bibinfo{person}{Pradeep Kuppala}, \bibinfo{person}{Ramona Comanescu}, \bibinfo{person}{Ramona Merhej}, \bibinfo{person}{Reena Jana}, \bibinfo{person}{Reza~Ardeshir Rokni}, \bibinfo{person}{Rishabh Agarwal}, \bibinfo{person}{Ryan Mullins}, \bibinfo{person}{Samaneh Saadat}, \bibinfo{person}{Sara~Mc Carthy}, \bibinfo{person}{Sarah Cogan}, \bibinfo{person}{Sarah Perrin}, \bibinfo{person}{Sébastien M.~R. Arnold}, \bibinfo{person}{Sebastian Krause}, \bibinfo{person}{Shengyang Dai}, \bibinfo{person}{Shruti Garg}, \bibinfo{person}{Shruti Sheth}, \bibinfo{person}{Sue Ronstrom}, \bibinfo{person}{Susan Chan}, \bibinfo{person}{Timothy Jordan}, \bibinfo{person}{Ting Yu}, \bibinfo{person}{Tom Eccles}, \bibinfo{person}{Tom Hennigan}, \bibinfo{person}{Tomas Kocisky}, \bibinfo{person}{Tulsee Doshi}, \bibinfo{person}{Vihan Jain}, \bibinfo{person}{Vikas Yadav}, \bibinfo{person}{Vilobh Meshram}, \bibinfo{person}{Vishal Dharmadhikari}, \bibinfo{person}{Warren Barkley}, \bibinfo{person}{Wei Wei}, \bibinfo{person}{Wenming Ye}, \bibinfo{person}{Woohyun Han}, \bibinfo{person}{Woosuk Kwon}, \bibinfo{person}{Xiang Xu}, \bibinfo{person}{Zhe Shen}, \bibinfo{person}{Zhitao Gong}, \bibinfo{person}{Zichuan Wei}, \bibinfo{person}{Victor Cotruta}, \bibinfo{person}{Phoebe Kirk}, \bibinfo{person}{Anand Rao}, \bibinfo{person}{Minh Giang}, \bibinfo{person}{Ludovic Peran}, \bibinfo{person}{Tris Warkentin}, \bibinfo{person}{Eli Collins}, \bibinfo{person}{Joelle Barral}, \bibinfo{person}{Zoubin Ghahramani}, \bibinfo{person}{Raia Hadsell}, \bibinfo{person}{D. Sculley}, \bibinfo{person}{Jeanine Banks}, \bibinfo{person}{Anca Dragan}, \bibinfo{person}{Slav Petrov}, \bibinfo{person}{Oriol Vinyals}, \bibinfo{person}{Jeff Dean}, \bibinfo{person}{Demis Hassabis}, \bibinfo{person}{Koray Kavukcuoglu}, \bibinfo{person}{Clement Farabet}, \bibinfo{person}{Elena Buchatskaya}, \bibinfo{person}{Sebastian Borgeaud}, \bibinfo{person}{Noah Fiedel}, \bibinfo{person}{Armand Joulin}, \bibinfo{person}{Kathleen Kenealy}, \bibinfo{person}{Robert Dadashi}, {and} \bibinfo{person}{Alek Andreev}.} \bibinfo{year}{2024}\natexlab{}.
\newblock \bibinfo{title}{Gemma 2: Improving Open Language Models at a Practical Size}.
\newblock
\showeprint[arxiv]{2408.00118}~[cs.CL]
\urldef\tempurl%
\url{https://arxiv.org/abs/2408.00118}
\showURL{%
\tempurl}


\bibitem[Vaswani et~al\mbox{.}(2017)]%
        {attention-is-all-you-need}
\bibfield{author}{\bibinfo{person}{Ashish Vaswani}, \bibinfo{person}{Noam Shazeer}, \bibinfo{person}{Niki Parmar}, \bibinfo{person}{Jakob Uszkoreit}, \bibinfo{person}{Llion Jones}, \bibinfo{person}{Aidan~N. Gomez}, \bibinfo{person}{\L{}ukasz Kaiser}, {and} \bibinfo{person}{Illia Polosukhin}.} \bibinfo{year}{2017}\natexlab{}.
\newblock \showarticletitle{Attention is all you need}. In \bibinfo{booktitle}{\emph{Proceedings of the 31st International Conference on Neural Information Processing Systems}} (Long Beach, California, USA) \emph{(\bibinfo{series}{NIPS'17})}. \bibinfo{publisher}{Curran Associates Inc.}, \bibinfo{address}{Red Hook, NY, USA}, \bibinfo{pages}{6000–6010}.
\newblock
\showISBNx{9781510860964}


\bibitem[Wang et~al\mbox{.}(2021)]%
        {autosa}
\bibfield{author}{\bibinfo{person}{Jie Wang}, \bibinfo{person}{Licheng Guo}, {and} \bibinfo{person}{Jason Cong}.} \bibinfo{year}{2021}\natexlab{}.
\newblock \showarticletitle{AutoSA: A Polyhedral Compiler for High-Performance Systolic Arrays on FPGA}. In \bibinfo{booktitle}{\emph{The 2021 ACM/SIGDA International Symposium on Field-Programmable Gate Arrays}} (Virtual Event, USA) \emph{(\bibinfo{series}{FPGA '21})}. \bibinfo{publisher}{Association for Computing Machinery}, \bibinfo{address}{New York, NY, USA}, \bibinfo{pages}{93–104}.
\newblock
\showISBNx{9781450382182}
\href{https://doi.org/10.1145/3431920.3439292}{doi:\nolinkurl{10.1145/3431920.3439292}}


\bibitem[Wang et~al\mbox{.}(2025)]%
        {COSAPlus}
\bibfield{author}{\bibinfo{person}{Zhican Wang}, \bibinfo{person}{Gang Wang}, {and} \bibinfo{person}{Guanghui He}.} \bibinfo{year}{2025}\natexlab{}.
\newblock \showarticletitle{COSA Plus: Enhanced Co-Operative Systolic Arrays for Attention Mechanism in Transformers}.
\newblock \bibinfo{journal}{\emph{IEEE Transactions on Computer-Aided Design of Integrated Circuits and Systems}} \bibinfo{volume}{44}, \bibinfo{number}{2} (\bibinfo{year}{2025}), \bibinfo{pages}{723--736}.
\newblock
\href{https://doi.org/10.1109/TCAD.2024.3434447}{doi:\nolinkurl{10.1109/TCAD.2024.3434447}}


\bibitem[Wang et~al\mbox{.}(2023)]%
        {COSA}
\bibfield{author}{\bibinfo{person}{Zhican Wang}, \bibinfo{person}{Gang Wang}, \bibinfo{person}{Honglan Jiang}, \bibinfo{person}{Ningyi Xu}, {and} \bibinfo{person}{Guanghui He}.} \bibinfo{year}{2023}\natexlab{}.
\newblock \showarticletitle{COSA:Co-Operative Systolic Arrays for Multi-head Attention Mechanism in Neural Network using Hybrid Data Reuse and Fusion Methodologies}. In \bibinfo{booktitle}{\emph{2023 60th ACM/IEEE Design Automation Conference (DAC)}}. \bibinfo{pages}{1--6}.
\newblock
\href{https://doi.org/10.1109/DAC56929.2023.10247678}{doi:\nolinkurl{10.1109/DAC56929.2023.10247678}}


\bibitem[Weng et~al\mbox{.}(2020)]%
        {dsagen}
\bibfield{author}{\bibinfo{person}{Jian Weng}, \bibinfo{person}{Sihao Liu}, \bibinfo{person}{Vidushi Dadu}, \bibinfo{person}{Zhengrong Wang}, \bibinfo{person}{Preyas Shah}, {and} \bibinfo{person}{Tony Nowatzki}.} \bibinfo{year}{2020}\natexlab{}.
\newblock \showarticletitle{DSAGEN: Synthesizing Programmable Spatial Accelerators}. In \bibinfo{booktitle}{\emph{2020 ACM/IEEE 47th Annual International Symposium on Computer Architecture (ISCA)}}. \bibinfo{pages}{268--281}.
\newblock
\href{https://doi.org/10.1109/ISCA45697.2020.00032}{doi:\nolinkurl{10.1109/ISCA45697.2020.00032}}


\bibitem[Wilson(2024)]%
        {verilator}
\bibfield{author}{\bibinfo{person}{C. Wilson}.} \bibinfo{year}{2024}\natexlab{}.
\newblock \bibinfo{title}{Verilator: The fastest free Verilog HDL simulator}.
\newblock \bibinfo{howpublished}{\url{https://verilator.org/}}.
\newblock
\newblock
\shownote{Version 5.024}.


\bibitem[Xu et~al\mbox{.}(2025)]%
        {mosaic}
\bibfield{author}{\bibinfo{person}{Jianxing Xu}, \bibinfo{person}{Yuanbo Wen}, \bibinfo{person}{Zikang Liu}, \bibinfo{person}{Ruibai Xu}, \bibinfo{person}{Tingfeng Ruan}, \bibinfo{person}{Jun Bi}, \bibinfo{person}{Rui Zhang}, \bibinfo{person}{Di Huang}, \bibinfo{person}{Xinkai Song}, \bibinfo{person}{Yifan Hao}, \bibinfo{person}{Xing Hu}, \bibinfo{person}{Zidong Du}, \bibinfo{person}{Chongqing Zhao}, \bibinfo{person}{Jiang Jie}, {and} \bibinfo{person}{Qi Guo}.} \bibinfo{year}{2025}\natexlab{}.
\newblock \showarticletitle{Mosaic: Exploiting Instruction-Level Parallelism on Deep Learning Accelerators with iTex Tessellation}. In \bibinfo{booktitle}{\emph{Proceedings of the 30th ACM International Conference on Architectural Support for Programming Languages and Operating Systems, Volume 2}} (Rotterdam, Netherlands) \emph{(\bibinfo{series}{ASPLOS '25})}. \bibinfo{publisher}{Association for Computing Machinery}, \bibinfo{address}{New York, NY, USA}, \bibinfo{pages}{672–688}.
\newblock
\showISBNx{9798400710797}
\href{https://doi.org/10.1145/3676641.3716262}{doi:\nolinkurl{10.1145/3676641.3716262}}


\bibitem[Yang et~al\mbox{.}(2025)]%
        {qwen25}
\bibfield{author}{\bibinfo{person}{An Yang}, \bibinfo{person}{Baosong Yang}, \bibinfo{person}{Beichen Zhang}, \bibinfo{person}{Binyuan Hui}, \bibinfo{person}{Bo Zheng}, \bibinfo{person}{Bowen Yu}, \bibinfo{person}{Chengyuan Li}, \bibinfo{person}{Dayiheng Liu}, \bibinfo{person}{Fei Huang}, \bibinfo{person}{Haoran Wei}, \bibinfo{person}{Huan Lin}, \bibinfo{person}{Jian Yang}, \bibinfo{person}{Jianhong Tu}, \bibinfo{person}{Jianwei Zhang}, \bibinfo{person}{Jianxin Yang}, \bibinfo{person}{Jiaxi Yang}, \bibinfo{person}{Jingren Zhou}, \bibinfo{person}{Junyang Lin}, \bibinfo{person}{Kai Dang}, \bibinfo{person}{Keming Lu}, \bibinfo{person}{Keqin Bao}, \bibinfo{person}{Kexin Yang}, \bibinfo{person}{Le Yu}, \bibinfo{person}{Mei Li}, \bibinfo{person}{Mingfeng Xue}, \bibinfo{person}{Pei Zhang}, \bibinfo{person}{Qin Zhu}, \bibinfo{person}{Rui Men}, \bibinfo{person}{Runji Lin}, \bibinfo{person}{Tianhao Li}, \bibinfo{person}{Tianyi Tang}, \bibinfo{person}{Tingyu Xia}, \bibinfo{person}{Xingzhang Ren}, \bibinfo{person}{Xuancheng Ren}, \bibinfo{person}{Yang Fan}, \bibinfo{person}{Yang Su}, \bibinfo{person}{Yichang Zhang}, \bibinfo{person}{Yu Wan}, \bibinfo{person}{Yuqiong Liu}, \bibinfo{person}{Zeyu Cui}, \bibinfo{person}{Zhenru Zhang}, {and} \bibinfo{person}{Zihan Qiu}.} \bibinfo{year}{2025}\natexlab{}.
\newblock \bibinfo{title}{Qwen2.5 Technical Report}.
\newblock
\showeprint[arxiv]{2412.15115}~[cs.CL]
\urldef\tempurl%
\url{https://arxiv.org/abs/2412.15115}
\showURL{%
\tempurl}


\bibitem[Yang et~al\mbox{.}(2020)]%
        {interstellar}
\bibfield{author}{\bibinfo{person}{Xuan Yang}, \bibinfo{person}{Mingyu Gao}, \bibinfo{person}{Qiaoyi Liu}, \bibinfo{person}{Jeff Setter}, \bibinfo{person}{Jing Pu}, \bibinfo{person}{Ankita Nayak}, \bibinfo{person}{Steven Bell}, \bibinfo{person}{Kaidi Cao}, \bibinfo{person}{Heonjae Ha}, \bibinfo{person}{Priyanka Raina}, \bibinfo{person}{Christos Kozyrakis}, {and} \bibinfo{person}{Mark Horowitz}.} \bibinfo{year}{2020}\natexlab{}.
\newblock \showarticletitle{Interstellar: Using Halide's Scheduling Language to Analyze DNN Accelerators}. In \bibinfo{booktitle}{\emph{Proceedings of the Twenty-Fifth International Conference on Architectural Support for Programming Languages and Operating Systems}} (Lausanne, Switzerland) \emph{(\bibinfo{series}{ASPLOS '20})}. \bibinfo{publisher}{Association for Computing Machinery}, \bibinfo{address}{New York, NY, USA}, \bibinfo{pages}{369–383}.
\newblock
\showISBNx{9781450371025}
\href{https://doi.org/10.1145/3373376.3378514}{doi:\nolinkurl{10.1145/3373376.3378514}}


\end{thebibliography}

\end{document}